\documentclass[aip,jcp,reprint,superscriptaddress,floatfix]{revtex4-2}
\def\SM{Supplementary Materials}

\usepackage{amsmath}
\usepackage{amssymb}
\usepackage{graphicx} 
\usepackage{booktabs}
\usepackage{comment}
\usepackage{braket}
\usepackage[english]{babel}
\usepackage{hyperref}

\usepackage[version=4]{mhchem}

\usepackage[usenames,dvipsnames]{xcolor}
\usepackage{soul}

\begin{document}
\include{defs}
\renewcommand{\vec}{\mathbf}

\title{Graph-neural-network predictions of solid-state NMR parameters from spherical tensor decomposition}

\author{Chiheb Ben Mahmoud}
\email{chiheb.benmahmoud@chem.ox.ac.uk}
\author{Louise A. M. Rosset}
\affiliation{Inorganic Chemistry Laboratory, Department of Chemistry, University of Oxford, Oxford OX1 3QR, United Kingdom}

\author{Jonathan R. Yates}
\affiliation{Department of Materials, University of Oxford, Oxford OX1 3PH, United Kingdom}

\author{Volker L. Deringer}
\affiliation{Inorganic Chemistry Laboratory, Department of Chemistry, University of Oxford, Oxford OX1 3QR, United Kingdom}

\date{\today}

\begin{abstract}
    Nuclear magnetic resonance (NMR) is a powerful spectroscopic technique that is sensitive to the local atomic structure of matter. Computational predictions of NMR parameters can help to interpret experimental data and validate structural models, and machine learning (ML) has emerged as an efficient route to making such predictions.
    Here, we systematically study graph-neural-network approaches to representing and learning tensor quantities for solid-state NMR -- specifically, the anisotropic magnetic shielding and the electric field gradient. 
    We assess how the numerical accuracy of different ML models translates into prediction quality for experimentally relevant NMR properties: chemical shifts, quadrupolar coupling constants, tensor orientations, and even static 1D spectra. 
    We apply these ML models to a structurally diverse dataset of amorphous \ce{SiO2} configurations, spanning a wide range of density and local order, to larger configurations beyond the reach of traditional first-principles methods, and to the dynamics of the $\alpha$--$\beta$ inversion in cristobalite.
    Our work marks a step toward streamlining ML-driven NMR predictions for both static and dynamic behavior of complex materials, and toward bridging the gap between first-principles modeling and real-world experimental data.
    
\end{abstract}

\maketitle

\section{Introduction}

Nuclear magnetic resonance (NMR) is widely used to probe the local atomic structure of materials and molecules~\cite{reif_solid-state_2021}, making it suitable to study static~\cite{pickard_first-principles_2002,holmes_structure_2024,dawson_site-directed_2024} and dynamic~\cite{spearing_dynamics_1992,sattig_nmr_2014,huang_27_2022} properties of chemical systems. Several interactions affecting nuclear spins are orientation-dependent, but they are usually averaged due to rapid tumbling, e.g., in liquid-state NMR on molecules. Therefore, isotropic parameters are often sufficient for most NMR-based analyses. In contrast, in solid-state materials, atoms are in well-defined geometric environments: even if experimental setups such as magic angle spinning can (partially) eliminate anisotropic interactions~\cite{polenova_magic_2015}, one still needs to include anisotropy to fully and rigorously interpret most solid-state NMR experiments.

The accurate computational prediction of (solid-state) NMR properties is therefore essential to interpret experimental data~\cite{ashbrook_combining_2016,valenzuela_reina_efg_2024} and could even be used in the design of materials by validating hypothetical structural models\cite{harper_modelling_2023}. First-principles computational methods based on density functional theory (DFT), especially the gauge-including projector augmented wave (GIPAW) method~\cite{pickard_all-electron_2001,yates_calculation_2007}, provide accurate predictions of NMR parameters for extended systems. Yet, these methods are still computationally expensive even on modern computing systems, limiting the time and length scales of problems that can be investigated from first principles. The high computational cost quickly becomes prohibitive when targeting structurally complex, e.g., disordered configurations, or for high-throughput screening.

Machine-learning (ML) methods have now been well established to circumvent the limitations of ``classical'' first-principles modeling. In particular, ML-based interatomic potentials (MLIPs)~\cite{behler_generalized_2007,bartok_gaussian_2010,zhang_deep_2018,drautz_atomic_2019,schutt_schnetpack_2019} have been the cornerstone of accelerating atomistic modeling while maintaining first-principles accuracy~\cite{li_molecular_2015}. These ML-driven simulations, as reviewed, for example, in Refs.~\citenum{unke_machine_2021} and \citenum{deringer_machine_2019}, have provided insights into, say, the mechanisms of structural transitions~\cite{cheng_evidence_2020,deringer_origins_2021,deng_machine_2023} and nucleation processes~\cite{sosso_fast_2013,piaggi_homogeneous_2022}.
Beyond MLIPs, atomistic ML methods have been applied to learn and predict a wide range of properties in materials and molecules: scalar quantities such as ionization energies~\cite{rupp_fast_2012}, band gaps~\cite{talapatra_band_2023}, and heat capacities~\cite{isayev_universal_2017}, as well as tensorial properties such as the dielectric response~\cite{wilkins_accurate_2019,grumet_delta_2024} and polarization~\cite{gigli_thermodynamics_2022}. 
In addition, ML techniques enable predictions of electronic-structure properties~\cite{grisafi_transferable_2019,chandrasekaran_solving_2019,ben_mahmoud_learning_2020,dey_machine_2023,fiedler_predicting_2023,scherbela_towards_2024}, and extend to spectroscopic fingerprints such as X-ray photoelectron spectroscopy~\cite{sun_machine_2022,golze_accurate_2022} 
to further bridge the gap between atomistic modeling and experimental observables.

Recently, geometric equivariant graph neural networks (GNNs) have emerged as powerful tools in atomistic modeling, providing cost-effective integration of chemical information. These architectures are especially promising for building ML models targeting a wide range of atomic species and properties. Notably, they enabled the fitting of MLIP models applicable to many elements across the Periodic Table~\cite{batatia_foundation_2023,deng_chgnet_2023,merchant_scaling_2023,yang_mattersim_2024}.
A key advantage of GNNs lies in their incorporation of learnable atomic representations that respect symmetry requirements under geometric transformations, such as rotations. This structural feature makes them particularly well-suited for modeling a wide range of NMR parameters, such as magnetic shielding (MS) and electric field gradient (EFG). Early applications of GNNs to learn NMR parameters focused on the isotropic contribution to the MS in proteins and glassy materials~\cite{kwon_neural_2020,han_scalable_2022,bankestad_carbohydrate_2024}. These models define the current state-of-the-art accuracy compared to earlier ML models leveraging fixed atomic representations~\cite{rupp_machine_2015,paruzzo_chemical_2018,chaker_nmr_2019}. 
By embedding physical symmetries within their architecture, GNNs are not only effective for isotropic targets but also extend to anisotropic NMR parameters, broadening their applicability across diverse molecular and material systems. In particular, Venetos and collaborators~\cite{venetos_machine_2023} found that learning the MS tensor and then extracting the anisotropy parameters is more accurate than developing specialized ML models for these quantities. This result was also confirmed in other works using fixed atomic representations in predicting the MS~\cite{charpentier_first-principles_2024} and EFG~\cite{harper_performance_2024} tensors.

Here, we build on these advances and present an ML framework aimed at predicting tensorial NMR properties, specifically magnetic shielding and the electric field gradient, using the NequIP~\cite{batzner_e3-equivariant_2022} GNN architecture. We explore two strategies for decomposing and representing NMR tensors as targets of ML models leveraging higher-rank tensors, assessing their predictive performance for both isotropic and anisotropic properties, such as the skewness parameters and the orientation of the NMR tensors. 
We benchmark our methods on a structurally diverse dataset of amorphous silica (a-\ce{SiO2}) structures obtained at various quenching rates. 
We demonstrate the utility of our ML tensorial models by constructing static one-dimensional (1D) NMR spectra of structural models of a-\ce{SiO2} at varying densities and of hypothetical zeolites, as well as the high-temperature $\alpha$--$\beta$ inversion in cristobalite. Our findings highlight the potential of combining ML-based models to accurately explore the potential energy surface and to predict anisotropic NMR properties in complex material systems. 
\section{Methods}

\subsection{NMR parameters}\label{sec:nmr_params}

The general form of a term in the NMR interaction Hamiltonian~\cite{haeberlen_high_1976}, which links the nuclear spin vector angular momentum operator $\vec{I}$ to an interacting (external or internal) vector field $\vec{S}^\lambda$, can be expressed as
\begin{equation*}
    H^\lambda = c^\lambda\cdot \vec{I}\cdot \vec{T}^\lambda\cdot \vec{S}^\lambda,
\end{equation*} 
where $c^\lambda$ is an interaction-specific constant, and $\vec{T}^\lambda$ is a rank-2 Cartesian tensor describing the response of the nuclear spin to $\vec{S}^\lambda$. 
For example, the magnetic shielding (MS) interaction parameter $\boldsymbol{\sigma}$ links a nuclear-spin operator $\vec{I}$ to an external magnetic field $\vec{B}_0$: 
$$H^{\mathrm{MS}}=\gamma^{\mathrm{MS}}~\vec{I}\cdot \boldsymbol{\sigma}\cdot \vec{B}_0,$$ 
where $\gamma^{\mathrm{MS}}$ is the gyromagnetic ratio. The MS parameter describes the response of the electrons, relating the induced magnetic field $\vec{B}_{\mathrm{ind}}$ to $\vec{B}_0$:
$$\vec{B}_{\mathrm{ind}} = -\boldsymbol{\sigma} \vec{B}_0.$$
Similarly, the electric field gradient (EFG) tensor $\vec{V}$ describes the interaction of a quadrupolar nucleus (nuclear spin $I~>~1/2$) with the electric field gradient produced by the electronic charge density. This  couples the spin operator $\vec{I}$ to itself, viz.
$$H^{Q} = c^Q~\vec{I}\cdot\mathbf{V}\cdot\vec{I},$$
where the constant $c^Q=\frac{eQ}{2I(2I-1)\hbar}$, $e$ is the electron elemental charge, $Q$ is the nuclear quadrupole moment, and $\hbar$ is the reduced Planck constant. The EFG tensor is traceless symmetric, and is usually expressed as the second spatial derivative of the electric field $V$ at the quadrupolar nucleus:
$$\vec{V} = \frac{\partial^2 V}{\partial x_i \partial x_j}.$$

In liquid-state systems, rapid tumbling means that only the isotropic averages of NMR tensor are needed to simulate experimental spectra. However, for solid-state systems, the full tensor nature of  $\boldsymbol{\sigma}$ and $\vec{V}$ must be considered. Usually, it is not practical to communicate all nine components of the rank-2 tensor directly, because these tensors transform under symmetry operations, such as rotations. Two NMR tensors might have different Cartesian coordinates, but they are linked by a rotation, therefore these two tensors represent the same NMR parameter. The NMR community has consequently developed several conventions~\cite{harris_further_2008} to report the MS tensor. The most common ones are the Maryland parameters~\cite{mason_conventions_1993}: the span $\Omega$ and the skew $\kappa$; and the Haeberlen parameters~\cite{noauthor_advances_1976}: the anisotropy $\zeta$ and the asymmetry $\eta$. 
These conventions capture the anisotropic nature of $\boldsymbol{\sigma}$ from its three eigenvalues. In the \SM, we summarize the analytical expression of these quantities. These sets of conventions simplify reporting and comparison between systems; hence their use in various software packages for spin dynamics simulations. 
These conventions can be extended to the description of the EFG tensor. In what follows, we use the subscript ``$\sigma$'' to refer to quantities derived from the MS tensor, and the subscript ``$Q$'' to refer to quantities derived from the EFG tensor.

\subsection{Representations of NMR tensors as spherical tensors}\label{sec:tens_decom}
\begin{figure*}
    \centering
    \includegraphics[]{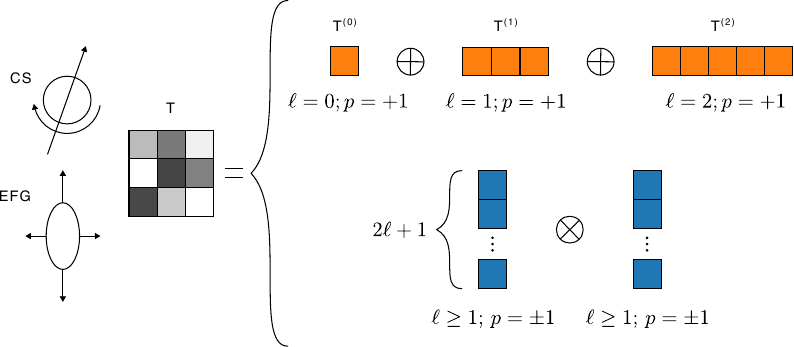}
    \caption{Spherical decomposition strategies for a rank-2 spherical tensor, denoted $\mathbf{T}$, as ML algorithm targets. Orange: irreducible spherical decomposition. Blue: the tensor product of rank $\ell$ tensors.}
    \label{fig:tensor_decomp}
\end{figure*}

As discussed above, the Cartesian representation is not ideal to represent NMR parameters because the coordinates can undergo geometric transformations. This makes the construction of ML models for the Cartesian tensors challenging, as one might introduce two nearly-identical atomic environments in the training set with substantially different tensorial NMR parameters, but where these two parameters are related by a rotation. This would make the learning process less robust, as identical data points have different labels. We hence choose to transform the NMR parameters into spherical tensors, as they have a unique decomposition (up to permutation). This choice allows us to leverage the capabilities of modern geometric GNN architectures, such as NequIP, which utilize spherical harmonics to achieve equivariance. 
Our choice is also justified by the implementation of Ref.~\citenum{venetos_machine_2023} based on the Tensor Field Networks architecture~\cite{thomas_tensor_2018}. The authors found that transforming the Cartesian NMR tensors into spherical tensors yielded the best prediction errors for the MS tensors of silicon. The spherical transformation enables us to apply the angular momentum coupling formalism developed in quantum mechanics (QM), allowing for an efficient decomposition of the NMR tensors and capturing their rotational properties. 

In the present work, we explore two different approaches to decomposing the rank-2 NMR tensors $\mathbf{T}$ as the learning target of an equivariant ML model. Figure~\ref{fig:tensor_decomp} provides a graphical summary of the two methods. The first and straightforward approach involves projecting $\mathbf{T}$ into the natural basis of rotations, i.e., spherical harmonics. We refer to this approach as the irreducible spherical decomposition (ISD). It leads to decomposing $\mathbf{T}$ as follows:
\begin{equation}\label{eq:isd}
    \mathbf{T} = T^{(0)} \oplus T^{(1)} \oplus T^{(2)} ,
\end{equation}
where $T^{(0)},~ T^{(1)},~ T^{(2)}$ are the spherical components corresponding to ranks $\ell=0$, $1$, and $2$, with dimensions 1, 3, and 5, respectively. This decomposition allows each component $T^{(i)}$ to transform independently, hence they can be targeted by independent ML models. 
In addition to their rotational properties, we should also consider the transformation of the $T^{(i)}$ under the parity operator. All the components $T^{(0)}$, $T^{(1)}$, and $T^{(2)}$ exhibit \textit{even} parity, i.e. their coordinates are invariant under the transformation reflecting their coordinates. This means that $T^{(1)}$ transforms like a pseudovector, unlike ``regular'' vectors, which exhibit odd parity behavior. $T^{(0)}$ and $T^{(2)}$ can be mapped to the symmetric part of $T$, while $T^{(1)}$ can be mapped to its antisymmetric part. 

Applying the ISD to the MS tensors yields a decomposition of $\boldsymbol{\sigma}$ into 3 non-zero irreducible components: $\sigma^{(0)}$, $\sigma^{(1)}$, and $\sigma^{(2)}$, corresponding to the ranks $\ell=0,~1,~\mathrm{and}~2$, respectively. The symmetric parts of the MS tensor dominate the experimentally observed NMR signal. The remaining antisymmetric part of the tensor, i.e. the $\ell=1$ contribution, has only a second-order effect on the NMR spectra. It therefore has minimal impact on the Zeeman transitions~\cite{bonhomme_first-principles_2012}, but there is some evidence of its role in relaxation processes~\cite{anet_nmr_1990}. The EFG tensor is symmetric and traceless; hence it can be fully described by $V^{(2)}$, which is the $\ell=2$ contribution in Eq.~\eqref{eq:isd}.

The second approach to decompose anisotropic tensors leverages a property of the tensor product of two angular momentum operators of ranks $l_1$ and $l_2$. In the context of angular momentum coupling, the tensor product $T^{(l_1)} \otimes T^{(l_2)}$ can be decomposed into a direct sum of irreducible representations corresponding to ranks ranging from $|l_1 - l_2|$ to $l_1 + l_2$:
\begin{equation}\label{eq:tensor_product}
    T^{l_1} \otimes T^{l_2} = \bigoplus_{l = |l_1 - l_2|}^{l_1 + l_2} T^{(l)}.
\end{equation}
This property allows us to systematically break down complex tensor interactions into simpler, well-defined components, which can be more effectively targeted by an equivariant ML model. By setting specific values for $l_1\geq1$ and $l_2\geq1$, we can recover the ISD decomposition of a rank-2 Cartesian tensor as expressed in Eq.~\eqref{eq:isd}. In particular, obtaining the rank $\ell=0$ tensor requires enforcing the equality $l_1=l_2$. Similar arguments can be made for the ranks $\ell=1$ and $\ell=2$. It is worth noting that the parity of both tensors $T^{(l_1)}$ and $T^{(l_2)}$ must match -- otherwise, the decomposition will yield irreducible tensors with the wrong parity. 
Finally, to further increase the flexibility of the decomposition of Eq.~\eqref{eq:tensor_product}, one can use a linear combination, with possibly learnable weights, of $N^2$ tensor products:
\begin{equation}\label{eq:tensor_product_expansion}
    \sum_{k_1=1}^N \sum_{k_2=1}^N w_{k_1 k_2} , T_{k_1}^{(l_1)} \otimes T_{k_2}^{(l_2)} = \bigoplus_{l = |l_1 - l_2|}^{l_1 + l_2} T^{(l)},
\end{equation}
where $w_{k_1k_2}$ are the linear expansion coefficients.

\subsection{Implementation in NequIP}
We implement the tensor decompositions detailed in Eqs.~\eqref{eq:isd}--\eqref{eq:tensor_product_expansion} within the widely used NequIP GNN architecture. This choice is motivated by NequIP's use of an atom-centered decomposition of the global property of a configuration, which allow us to define per-atom learning targets. Additionally, it employs spherical harmonics with both even and odd parities to achieve equivariance in its ML models. This enables us to map the collection of spherical tensors used to reconstruct an NMR parameter to those of internal representation of the ML model.

NequIP, as in many other MLIPs~\cite{behler_generalized_2007, bartok_gaussian_2010, drautz_atomic_2019}, aims to model a global (scalar) target $A$, such as the total energy $E$ of a structure, by decomposing it into a sum of atomic contributions $A_i$: $A = \sum_i A_i$. The NequIP architecture models the $A_i$ from a set of features, including the chemical species of the central atom and its neighbors and the distances between neighbors, which correspond to a radial description of the system. 
The angular information is captured using spherical harmonics with trainable weights. These features are combined and updated at every (interaction) layer of the neural-network architecture via a convolution. 
At each layer, the output is aggregated with similar messages from neighboring atoms and then passed to the next layer. 
Unlike the typical NequIP energy model which discards the non-invariant angular features (i.e. those with $\ell>0$), at the last layer, we map these components to the irreducible tensors of interest through an \texttt{e3nn}~\cite{geiger_e3nn_2022} geometric linear layer.

For the ISD, the output is straightforward and corresponds to the irreducible tensor $T^{(i)}$. For the tensor product representation, we obtain two sets of $N$ tensors of rank $\ell\geq 1$, and then calculate the tensor products between them, resulting in a total of $N^2$ tensor products. 
At this stage, we only keep the $\ell\leq2$ components as, according to Eq.~\eqref{eq:isd}, they are the only useful ones to reconstruct the full NMR tensors. 
Notably, this approach also allows for targeting a single irreducible tensor of rank $\ell$, which can be useful if an ML architecture does not explicitly incorporate certain combinations of spherical tensor rank and parity (for example, $\ell=1$ and even parity which corresponds to a pseudovector). Finally, we skip the summation over atomic contributions that would be used in MLIPs, because our output NMR parameters are all defined per atom.

\subsection{Datasets}
\begin{figure*}
    \centering
    \includegraphics[width=1\linewidth]{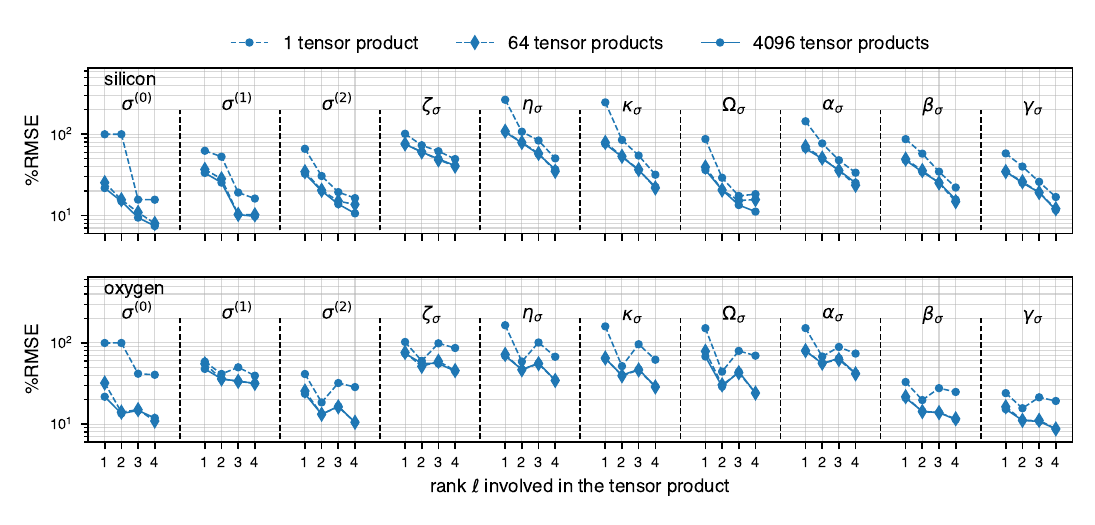}
    \caption{Evolution of ML model errors, reported on the isotropic and anisotropic properties of the MS tensor, as a function of the rank of the tensors involved in the tensor product. Dashed lines with circle markers refer to a single tensor product involved, the dashed lines with diamond markers to a learnable combination of 64 tensor products, and the solid lines to a learnable linear combination of 4,096 tensor products. Upper panel: silicon atoms, lower panel: oxygen atoms.}
    \label{fig:tensor_tp_decomp}
\end{figure*}
To train and evaluate our ML models, we construct a challenging dataset of 1,000 a-\ce{SiO2} structures obtained through melt--quench--anneal molecular dynamics (MD) simulations using LAMMPS~\cite{thompson_lammps_2022}, spanning a wide range of densities from $2.0$ to $2.7$~g~cm$^{-3}$. We follow the protocol of Ref.~\citenum{erhard_machine-learned_2022} to generate the a-\ce{SiO2} configurations, as it yields a good balance between accuracy and simulation speed. For each target density, we melt an initial configuration of $144$ atoms for $10$~ps at $3,500$~K, then quench to $300$~K, using the empirical interatomic potential by Carré, Horbach, Ispas, and Kob (CHIK)~\cite{carre_new_2008}. To maximize the diversity of local atomic environments, we use three quench rates: $10^{11},~ 10^{12},~\text{and }10^{13}~\mathrm{K~s}^{-1}$. Then, we use the Gaussian Approximation Potential (GAP) from Ref.~\citenum{erhard_machine-learned_2022} to anneal the resulting structures from the quenching simulations, within the $NVT$ and $NpT$ ensembles for $10$~ps. This protocol was shown to yield high-quality structural models of $a$-\ce{SiO2}.\cite{erhard_machine-learned_2022} In what follows, we use $800$ structures as training data for the ML models, $50$ structures for the internal validation and $150$ structures for testing. 

\subsection{Computational details}

We compute the NMR parameters, i.e., the MS and EFG tensors, using the GIPAW formalism as implemented in CASTEP~\cite{profeta_accurate_2003,clark_first_2005,yates_calculation_2007}. We use on-the-fly generated pseudopotentials and the Perdew--Burke--Ernzerhof (PBE)~\cite{perdew_generalized_1996} exchange--correlation functional. We use a fine $k$-point spacing of $0.03$~\AA$^{-1}$ with a grid offset of $(0.25,~0.25,~0.25)$, and a plane-wave energy cutoff of $900$~eV. We choose these parameters to ensure convergence of NMR tensors for both silicon and oxygen atoms. 

We determine the hyperparameters of the NequIP-based models using Bayesian optimization via the Cross-Platform Optimizer for Potentials (XPOT) package~\cite{thomas_du_toit_cross-platform_2023}. We randomly choose $100$ structures for this task, and select the best set of hyperparameters based on test errorsor $50$ independent configurations. For consistency, and because we have multiple representations for the NMR tensors, we only perform the optimization for a model targeting all elements in the irreducible spherical representation of the MS tensor [Eq.~\eqref{eq:isd}], and a model targeting $\ell=1$ in the case of the tensor product representation. We report the optimized hyperparameters in the \SM. In all of our models, we only train on normalized and standardized targets -- that is, we subtract the mean, and divide by the standard deviation of the quantity of interest across the training set. We then scale again by the standard deviation and add the mean at the inference step.

\section{Results and discussion}
Having discussed the details of modeling NMR tensors as ML targets, we now assess their performance on our a-\ce{SiO2} dataset. We begin by examining the effect of the number and the rank of the tensors involved in the tensor product on learning the MS tensors and anisotropy parameters from the tensor product representation. Next, we compare our best-performing tensor product model with the model resulting from the ISD. 

In our assessment of the performance of the ML models, we use the the root mean square error (RMSE): $$\mathrm{RMSE}=\sqrt{\frac{1}{N}\sum_{i=0}^N\left(y_i^{\mathrm{QM}} - y_i^{\mathrm{ML}}\right)^2},$$
and the normalized root mean square error (\%RMSE) defined as the RMSE divided by the standard deviation of the quantity of interest, expressed as a percentage:
$$ \%\mathrm{RMSE} = 100 \times \frac{\sqrt{\frac{1}{N}\sum_{i=0}^N\left(y_i^{\mathrm{QM}} - y_i^{\mathrm{ML}}\right)^2}}{\sqrt{\frac{1}{N}\sum_{i=0}^N\left(y_i^{\mathrm{QM}} - \bar{y}^{\mathrm{QM}}\right)^2}},$$
where $N$ is the number of data points in the test set, $y_i^{\mathrm{QM}}$ and $y_i^{\mathrm{ML}}$ are the QM target and the ML prediction, respectively, of data location $i$, and $\bar{y}^{\mathrm{QM}}$ is the average QM target in the dataset.
The use of \%RMSE, in particular, ensures a fair evaluation across multiple properties. In the following discussion, the error metrics are only provided for the testing set of 150 a-\ce{SiO2} structures. For a more comprehensive view of the performance of the ML models, we provide the RMSE values and the standard deviation (the spread) of the QM targets in the \SM.

Euler angles are usually computed as a set of three angle values describing the orientation of a rotation with respect to the reference axes. These three angles are not unique for any given rotation, as there are multiple sets of elemental rotations that combine to the same object~\cite{svenningsson_tensorview_2023}. In the next section, and due to their sensitivity, we treat each Euler angle separately. We compare the lowest values among the equivalent sets of angles. We use the $ZYZ$ ordering in the passive convention.

\subsection{Tensor product representation}

We begin by investigating the effect of two key ingredients of the tensor product representation: the rank $\ell$ of the tensors, and the number of the tensor products. First, we build four different models targeting a single tensor product, where the tensors involved have ranks $\ell$ ranging from 1 to 4. To ensure a fair comparison of model performance, we keep all other hyperparameters fixed, including the cutoff radius, the number of interaction layers, and the internal equivariant features with maximum rank $\ell_{\mathrm{max}}=4$. 
In Fig.~\ref{fig:tensor_tp_decomp}, we show, in dashed lines and circle markers, the evolution of the test set \%RMSE for both silicon and oxygen atoms and across the MS tensor properties discussed in Sec.~\ref{sec:nmr_params}. Our results show that the accuracy of ML models significantly increases with the rank $\ell$ of the tensors involved in the single tensor product. This improvement can be explained by the number of terms used in the tensor product of spherical harmonics. An element $T^{(\ell)} \equiv \ket{\ell,m}$, in the right hand side of Eq.~\eqref{eq:tensor_product}, in the bra--ket notations of QM, is given by~\cite{shankar_principles_2013}:
\begin{equation*}
    \ket{\ell,m} =  \sum_{m_1=-l}^{l} \sum_{m_2=-l}^{l} C_{lm_1lm_2}^{\ell m} \big(\ket{l;m_1} \otimes \ket{l;m_2}\big),
\end{equation*}
with $C_{lm_1lm_2}^{lm}$ being the Clebsch--Gordan coefficients. 
This equation shows that the number of terms used to reconstruct $T^{\ell}$ depends critically on the rank $l$ of the tensors and the selection rules of the $C_{lm_1lm_2}^{\ell m}$.  As the rank $l$ increases, the number of contributing terms grows, providing greater flexibility in capturing the necessary information to describe the target tensor properties. This observation can partially explain the poor performance of our $\ell=1$ and $\ell=2$ models at recovering the invariant contribution to MS tensor. 
Another factor contributing to the poor performance could be the number of target properties used to construct the loss function. The $\sigma^{(0)}$ is a single value, while $\sigma^{(1)}$ and $\sigma^{(2)}$ represent eight terms. This discrepancy means that the loss function may be disproportionately influenced by the higher-order terms. This could be investigated using a weighted loss function, but developing and carefully testing such approaches is beyond the scope of this work.

Having established the importance of flexibility in reconstructing the irreducible terms of NMR tensors, we now assess another approach to introducing such flexibility based on Eq.~\eqref{eq:tensor_product_expansion}. Instead of using a single tensor product, we employ a linear combination of tensor products with trainable weights. This method allows the model to learn the optimal combination of tensor products that represents the target properties; hence increasing the predictive power of the ML model.  
In Fig.~\ref{fig:tensor_tp_decomp}, we show the evolution of test-set errors for models using 64 tensor products, in dashed lines and diamond markers, and 4,096 tensor products, in solid lines and circle markers, to decompose the NMR tensors as a function of the rank of the decomposition tensors involved. We find that the performance of the ML models increases as the number of tensor products increases, even though the performance uplift for using 4,096 tensor products seems to be minimal. 
We also observe that higher ranks $\ell$ lead to more accurate models, for both silicon and oxygen atoms across all MS tensor properties. 
Furthermore, the performance of the ML models associated with $\ell=1$ and $\ell=2$ is significantly enhanced when using multiple tensor products, and compatible with the best models from the single tensor product representation. These results highlight the importance of incorporating a flexible target decomposition when building a reliable method in general-rank spherical tensor learning.

\subsection{Comparing the two decompositions}
\begin{figure}
    \centering
    \includegraphics[width=1\linewidth]{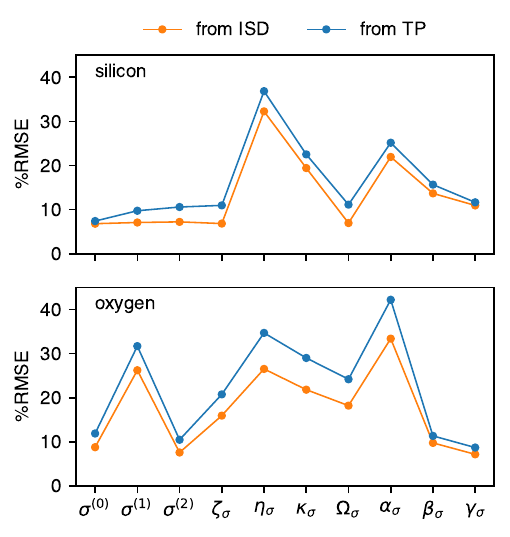}
    \caption{Comparison between the best models from the tensor product and the irreducible tensor decomposition (ISD) representations on the MS tensor of silicon and oxygen atoms in our a-\ce{SiO2} dataset. Errors are reported for the quantities discussed in Sec.~\ref{sec:nmr_params}.  ISD stands for irreducible tensor decomposition, and TP for tensor product, as discussed in Sec.~\ref{sec:tens_decom}. Upper panel: silicon atoms, lower panel: oxygen atoms. Orange: ISD representation, blue: tensor product representation.}
    \label{fig:sph_tp_test}
\end{figure}
We evaluate our best-performing model from the tensor-product representation against a direct model targeting ISD elements. Figure ~\ref{fig:sph_tp_test} summarizes our key findings. The two models exhibit no significant difference in the performance across all examined parameters of the MS tensors. We note that the tensor product model has $\approx 3$ times fewer trainable parameters, viz.\ 2,159,688 compared to 6,221,944 of the direct ISD model. This occurs because, despite using higher-rank hidden features for the tensor-product-based model, we are constrained to having fewer features per rank due to memory limitations. Nevertheless, our ML models reproduce the QM silicon MS tensors with excellent accuracy (below 20\% of the total variance of targets in the test set), except the asymmetry $\eta_\sigma$ and the span $\Omega_\sigma$. These parameters are obtained from $\sigma^{(2)}$ the traceless symmetric part of the MS tensor. This result highlights the sensitivity of these descriptors to the shape of the NMR tensor; hence the need for accurate predictions. 

\begin{figure}
    \centering
    \includegraphics[width=1\linewidth]{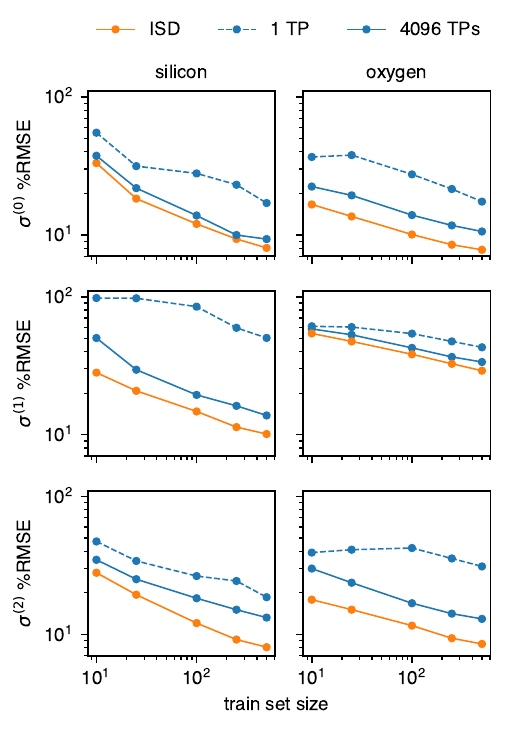}
    \caption{Learning curves for the irreducible tensors of the MS tensor for silicon atoms (left) and oxygen atoms (right). We only show the normalized errors \%RMSE of the test set. Each point is the average of $4$ partitions of the training set. ISD stands for irreducible tensor decomposition, and TP for tensor product, as discussed in Sec.~\ref{sec:tens_decom}. Dashed blue lines: single tensor product models; solid blue lines: 4,096 tensor product models; solid orange lines: ISD models.}
    \label{fig:lc}
\end{figure}

Our models' predictions for oxygen MS tensors are relatively less accurate than the silicon case. In particular, $\sigma^{(1)}$ is not well reproduced from both ML models compared to $\sigma^{(0)}$ and $\sigma^{(2)}$, with a \%RMSE of $26.14$\% and $31.70$\% corresponding to $0.80$ and $0.97$~ppm, for the ISD and tensor-product models respectively. Also, despite the good accuracy in predicting oxygen $\sigma^{(2)}$, all of the anisotropy parameters ($\eta_\sigma$, $\zeta_\sigma$, $\kappa_\sigma$, and $\Omega_\sigma$) show higher errors compared to the results for silicon. This can be explained by the wide range of oxygen MS parameters. 
These silicon and oxygen results highlight the need for more accurate metrics to characterize learning NMR parameters beyond the RMSE (and the mean absolute error) over the tensor components or the irreducible tensors. Harper and collaborators~\cite{harper_performance_2024} further explored this aspect in learning the EFG tensor. We believe that more systematic studies incorporating error metrics over a wide range of anisotropy parameters, either at the model evaluation step or at the model training step, could constitute a starting point for a future study.

To understand the limitations of our dataset in learning the MS tensors, we examine the learning curves (LCs) of all decomposition schemes discussed so far: ISD, $1$ tensor product, and $4096$ tensor products, as shown in Fig.~\ref{fig:lc}. The tensor product approaches are performed using the $\ell=4$ models. We evaluate prediction errors on the test set for the irreducible tensors using 4 folds of the training set. In all of our models, the LCs appear to be decreasing algebraically, except for the single tensor product model evaluated on the oxygen $\sigma^{(2)}$. This result indicates that the model accuracy can still be significantly improved by adding more training data. The oxygen $\sigma^{(1)}$ panel of Fig.~\ref{fig:lc} provides a partial explanation regarding the poor prediction errors of the oxygen $\sigma^{(1)}$, shown in Fig.~\ref{fig:sph_tp_test}. The slope of the LC of this label is not as steep compared to the other irreducible tensors. The \%RMSE is $\approx60\%$ for all tensor product models models when the training set size is $10$ structures, and only decreases to $\approx30\%$ for $500$ training structures in the case of the ISD model.

We also trained five models with different ``degrees of equivariance'' in their hidden features (cf. \SM), where the maximum rank $\ell_{\mathrm{max}}$ of the features ranges between 0 and 4. We found no significant improvement of the models' accuracy beyond $\ell_{\mathrm{max}}=2$, which is the minimum level of equivariance needed to learn the $\sigma^{(2)}$ contribution.

We study the EFG using only the ISD model. The EFG tensor is symmetric and traceless; hence it can be mapped directly to the $\ell=2$  in the irreducible spherical tensor decomposition (Eq.~\eqref{eq:isd}). In the \SM, we show ML prediction errors on the $\ell=2$ components, $V_{zz}$, which corresponds to the largest eigenvalue of the EFG, and is essential to describe quadrupolar interactions, as well as other anisotropic parameters of the EFG, similar to the MS tensor prediction assessment. In the early stages of this work, we verified that learning $V_{zz}$ with a specialized invariant model yielded worse prediction errors than an anisotropic model targeting the full EFG tensor. This result agrees with the findings in Ref.~\citenum{harper_performance_2024}.

\subsection{Simulated 1D NMR spectra}\label{sec:nmr_spectra}
\begin{figure}
    \centering
    \includegraphics[width=1\linewidth]{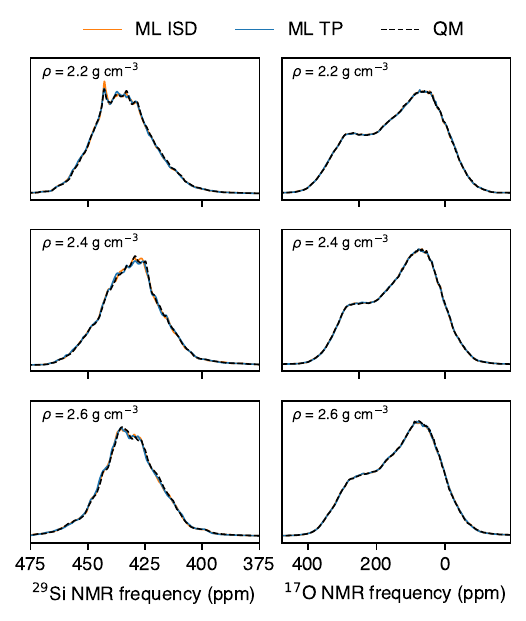}
    \caption{Simulated static 1D NMR spectra for three structural models of a-\ce{SiO2}, with densities of $2.2$, $2.4$, and $2.6$ g cm$^{-3}$ respectively. These models were obtained with a fast quench rate of $10^{13}$ K s$^{-1}$. Left column: $^{29}$\ce{Si} spectra, right column: $^{17}$\ce{O} spectra. Solid orange lines: spectra from ML prediction of NMR tensor parameters using the ISD; solid blue lines: spectra from ML prediction of NMR tensor parameters using the best tensor product model, dashed black lines: spectra from GIPAW-calculated NMR parameters.}
    \label{fig:spectra_density}
\end{figure}
Using the MLIP of Ref.~\citenum{erhard_modelling_2024} (because of the performance uplift compared to the GAP used to build the training set), we generated a series of a-\ce{SiO2} structural models at three different densities of $2.2,~2.4,~\text{and }2.6$ g cm$^{-3}$. Each structural model contains 300 atoms and was obtained using the same protocol as the training set at a quench rate of $10^{13}~\mathrm{K~s}^{-1}$ in the $NpT$ ensemble. This fast quench rate ensures diversity in the local structural order of the atomic environments. 
We use our ISD and best tensor-product models to predict the MS tensors, and the ISD model to predict the EFG tensors. Our models yield good accuracy, including for the problematic $\sigma^{(1)}$ for oxygen atoms. The full breakdown of errors is reported in the \SM. Then, we use SOPRANO\cite{noauthor_soprano_nodate} to simulate the static 1D NMR spectra from ML and GIPAW. Static spectra highlight the need for anisotropic NMR parameters, in contrast with other techniques like the magic angle spinning (MAS) that would eliminate the anisotropic effect. 
Figure~\ref{fig:spectra_density} shows the $^{29}\ce{Si}$ and $^{17}\ce{O}$ simulated spectra using a Gaussian broadening of $0.5$~ppm. Our ML spectra present good agreement with their GIPAW counterparts. The structural model of low density, $\rho=2.2$ g cm$^{-3}$, provides an abnormal a-\ce{SiO2} test case as it features a few large rings of sizes $20$ and $22$, alongside smaller rings with sizes ranging between $4$ and $18$. This distribution is markedly different from the training set, where most ring sizes are between $6$ and $16$, and the largest represented ring size is of 18 members, for a total training set of almost $61,000$ rings. 

\begin{figure}
    \centering
    \includegraphics[width=1\linewidth]{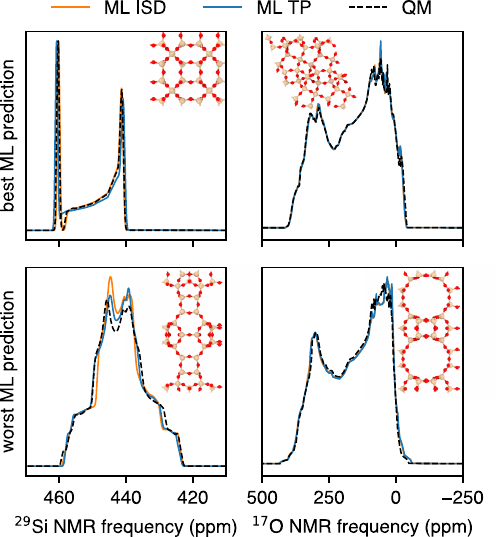}
    \caption{Best and worst ML-predicted simulated 1D static NMR spectra for four \ce{SiO2} hypothetical zeolites. The insets in each panel contain an illustration of the corresponding zeolite, rendered using OVITO\cite{stukowski_visualization_2010}. Left column: silicon spectra, right column: oxygen spectra. Solid orange lines: spectra from ML prediction using the ISD model of tensor NMR parameters; solid blue lines: spectra from ML prediction using the tensor product model of tensor NMR parameters; dashed black lines: spectra from QM calculated NMR parameters.
    }
    \label{fig:hypo_zeo}
\end{figure}

We assess the out-of-domain performance of our ML models on a dataset comprised of $50$ \ce{SiO2} hypothetical zeolites obtained from Ref.~\citenum{baerlocher_database_nodate}. In Fig.~\ref{fig:hypo_zeo}, we show the simulated static 1D $^{29}\ce{Si}$ and $^{17}\ce{O}$ NMR spectra of four zeolites obtained from the ISD and the best tensor product models, similar to the analysis on the a-\ce{SiO2} structural models. We also show the GIPAW-based spectra for comparison. We choose these four representative structures based on the accuracy of the ISD models in predicting the silicon MS and the oxygen MS and EFG tensors. We use the total error on the Cartesian tensors coordinates as our error metric for these evaluations. In the \SM, we report the errors on the irreducible contributions for the zeolites dataset. 
The tensor-product-based model outperforms the ISD model only for the isotropic MS values. This means that tensor-product models will have better predictions for the positions of the peaks, but not necessarily the shape of the spectra. This explains the slightly poorer quality of the spectra obtained from the tensor product model, compared to ISD model, in the best ML prediction silicon case where the anisotropic effects dominate the spectrum. 
However, the relatively higher accuracy of the isotropic MS yields a better spectrum where the anisotropic effects are less relevant, like in the case of the worst ML prediction of the $^{29}$\ce{Si} spectrum (bottom left panel of Fig.~\ref{fig:hypo_zeo}).
We find a possible correlation between the ring distribution in the zeolites and ML prediction errors. In particular, our best predicted spectra only had two ring sizes per element: $6$ and $12$ for silicon, and $8$ and $10$ for oxygen. In contrast, the worst predicted spectra correspond to zeolites with three sizes: $8$, $10$ and $12$ for both silicon and oxygen. These rings sizes are well represented in the training set. Further investigation of the possible influence of ring distributions is beyond the scope of the current work.

\begin{figure*}
    \centering
    \includegraphics[width=1\linewidth]{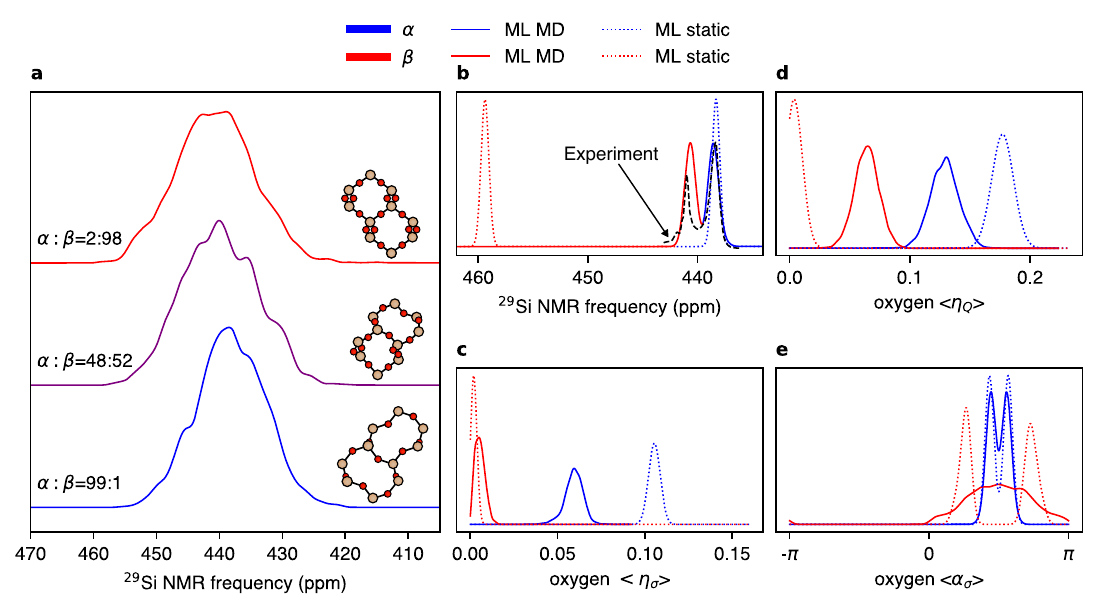}
    \caption{NMR fingerprints of an $\alpha$-cristobalite to $\beta$-cristobalite displacive phase transition. (a) Simulated $^{29}$\ce{Si} MAS spectra of a selection of MD snapshots describing a single $\alpha\rightarrow\beta$ transition. The $\alpha:\beta$ ratio reported on every spectrum describes the ratio of the $\alpha$-like and $\beta$-like environments in the configuration, obtained from the MLSI package~\cite{erhard_crystal_2024}. The small insets are an illustration of the typical environments found in the configuration. (b) The sum of $^{29}$\ce{Si} MAS spectra obtained for MS parameters averaged over the $\alpha$-like and $\beta$-like configurations. We compare to an experimentally obtained $^{29}$\ce{Si} powder MAS spectrum when both $\alpha$-cristobalite and $\beta$-cristobalite peaks were present in the spectrum, and to spectra of crystalline structures (data from Ref.~\citenum{spearing_dynamics_1992}). (c) The asymmetry of the averaged MS tensors over the $\alpha$-like and $\beta$-like configurations. (d) The asymmetry of the average EFG tensors over the same structure. (e) First Euler angle of the averaged MS tensors. Blue lines: $\alpha$-like behavior; red lines: $\beta$-like behavior. Solid lines: properties obtained from ML predictions over an average over the dynamics; dotted lines: properties obtained from ML predictions for crystalline structures; dashed line: experimental result. $<\cdot>$ denotes a quantity obtained from an averaged tensor over the trajectory.}
    \label{fig:cristo_inversion}
\end{figure*}

\section{Outlook: Dynamics of structural transitions}
In this section, we present one possible application of tensorial NMR ML models that goes beyond studying static structures. We investigate the dynamics of the \ce{SiO2} phase transition from $\alpha$- to $\beta$-cristobalite~\cite{walker_temperature_1958,peacor_high-temperature_1973}. We acknowledge that the crystal structure of $\beta$-cristobalite, in particular, is a controversial topic with multiple proposed structures~\cite{wright_structures_1975,hatch_alpha--beta_1991}. Here, we only aim to showcase of the usefulness of the tensorial NMR ML models, rather than reporting a full and detailed study or to resolve the structure as such.

We prepare an $\alpha$-cristobalite configuration containing 4,116 atoms, and run an $NpT$ MD trajectory, with $p=-2$~bar and $T=500$~K, for $100$~ps. We apply negative pressure to induce the volume expansion associated with the $\alpha$- to $\beta$-cristobalite inversion \cite{yuan__2012}. The MD simulation is driven by the same MLIP used to generate the structural models of Sec.~\ref{sec:nmr_spectra}. We use an ML-based classifier for \ce{SiO2} polymorphs~\cite{erhard_crystal_2024} to determine the percentage of $\alpha$-like and $\beta$-like environments in snapshots along the trajectory. The potential used in this work ``over-stabilizes'' the $\beta$ phase, as Erhard and collaborators argued in ~\citenum{erhard_machine-learned_2022}, presumably because it is trained on DFT data with the SCAN~\cite{sun_strongly_2015} exchange--correlation functional. As a result, we observe only a single transition event from the $\alpha$ to the $\beta$ phase, rather than multiple reversible transitions between $\alpha$ and $\beta$ phases in the same MD trajectory. We use our ISD ML models to predict the MS and EFG tensors along the trajectory. 

First, we characterize the $\alpha\rightarrow\beta$ transition using simulated $^{29}$\ce{Si} MAS spectra. We construct the MAS spectra from our ML predictions of the MS tensors for a selection of snapshots along the transition, as presented in Fig.~\ref{fig:cristo_inversion}(a). We show three spectra corresponding to (1) a configuration with predominantly $\alpha$ environments, (2) a transition configuration where the environments are equally distributed between the $\alpha$ and $\beta$ phases, and finally (3) a configuration with predominantly $\beta$ environments. The center of the distribution of the spectra shifts to higher frequencies as the percentage of $\alpha$ environments decreases and that of $\beta$ environments increases. 

Using our ML-driven methodology, we can take the average of NMR tensors over the $\alpha$- and $\beta$-labeled snapshots of the trajectory to create thermally-averaged tensors that capture the properties of both phases. A configuration is labeled $\alpha$ ($\beta$) if its $\alpha$ ($\beta$) environments constitute more than $90$~\% of the total environments. We then build the $^{29}$\ce{Si} MAS spectra from these two averaged tensors. In Fig.~\ref{fig:cristo_inversion}(b), we report the sum of these two signals, and compare to an experimentally obtained $^{29}$\ce{Si} MAS powder spectrum~\cite{spearing_dynamics_1992} for a cristobalite sample at $\approx500$~K, where half the sample is $\alpha$-cristobalite while the other half is $\beta$-cristobalite. Our ML predictions agree well with the experimental results. The separation between the experimental peaks, $\approx2.6$~ppm, is slightly larger that our ML-predicted one, $\approx2.3$~ppm. The experimental separation is even larger if we consider pure samples of $\alpha$- and $\beta$-cristobalite, and is measured to be $\approx4.2$~ppm. This discrepancy can be explained by the accuracy of the underlying GIPAW calculations with the PBE exchange--correlation functional. It may also suggest the need for a different averaging strategy. We may need longer trajectories, with multiple reversible transitions, to carry out averages with structures containing equal parts of $\alpha$ and $\beta$ phases. These investigations could constitute a study of its own.

We also compare, in Fig.~\ref{fig:cristo_inversion}(b), the thermally averaged spectra to ML-predicted static MAS spectra of crystalline $\alpha$- and $\beta$-cristobalite structures. The spectrum obtained for $\alpha$-cristobalite is in good agreement with the spectrum derived from the thermal average. However, the discrepancy in the position of the $\beta$-cristobalite peak is more pronounced, and estimated to be $\approx19$~ppm.
This suggests that thermal averaging may not recover the same crystalline properties at $0$~K, and caution must be taken when extrapolating from the average of the thermal motion of atoms due to possible finite temperature effects. This is a direction of further study.

The MAS spectra, discussed so far, only require isotropic MS contributions, corresponding to $\sigma^{(0)}$, which can be achieved using invariant (rather than equivariant) ML models. To leverage our anisotropic ML models, we look into the anisotropy parameters of the MS and EFG tensors. 
In particular, we report, in Fig.~\ref{fig:cristo_inversion}(c) and Fig.~\ref{fig:cristo_inversion}(d), the asymmetry of the oxygen atoms derived from the thermally averaged MS ($\eta_\sigma$) and EFG ($\eta_Q$) tensors for the $\alpha$ and $\beta$ phases. We also show the asymmetry obtained from the crystalline configurations. 
Then, we plot the parameters' distribution using kernel density estimation (KDE) analysis. 
We notice that the oxygen MS tensors for the (thermally averaged) $\alpha$-cristobalite simulation data are more skewed compared to $\beta$-cristobalite, despite both tensors exhibiting low anisotropic behavior. The asymmetry of the static crystals follows a similar trend. This trend persists also for the oxygen EFG tensors. In particular, our asymmetry parameter derived from the thermal averages ($\eta_Q=0.13$) is in excellent agreement with experimental results obtained by Pedro and Farnan and reported as unpublished in Ref.~\citenum{profeta_accurate_2003} ($\eta_Q=0.13$).

We further analyze these tensors by treating them as rotations. We extract the Euler angles of the averaged oxygen NMR tensors. In Fig.~\ref{fig:cristo_inversion}(e), we show the distribution of the Euler angle, associated with the averaged oxygen MS tensors, that exhibits a different distribution between the two phases, and compare it to the static crystalline limit. For the thermal averages, the Euler angle $\alpha_\sigma$ appears to have a bimodal distribution in $\alpha$-cristobalite, while it has a broad distribution in the $(0, ~\pi)$ interval in $\beta$-cristobalite. The crystalline limit only agrees with the dynamic average for the $\alpha$ phase, with a narrower distribution, while the crystalline $\beta$-cristobalite predicts a bimodal distribution centered around the edges of the $(0, ~\pi)$ interval. We presume that this broadening is an artifact of the thermal averaging, which could be further investigated with even larger cells and longer trajectories. We obtain overlapping distributions of the $\beta_\sigma$ and $\gamma_\sigma$ components for the two phases. These analyses are possible due to the ML-accelerated approach employed here, since we leverage the speed and accuracy of a ready-to-use MLIP for \ce{SiO2}, and couple it with fast and accurate ML models targeting anisotropic NMR parameters.

\section{Conclusions}
We have presented an ML framework based on the NequIP graph-neural-network architecture to predict tensorial NMR parameters, exploring two decomposition methods: (1) decomposing the NMR tensors into irreducible spherical components that can be targeted by independent models, and (2) leveraging angular momentum coupling to extract irreducible tensors from tensor products of identical-rank spherical tensors.
Testing on a diverse dataset of a-\ce{SiO2} configurations, we found that the irreducible spherical decomposition yielded the lowest prediction errors for the tensors and their derived properties, such the anisotropy parameters and tensor orientations. The tensor-product-based approach achieved similar accuracy either by using single high-rank tensor products or a trainable linear combination of tensor products. Both models produce simulated static 1D NMR spectra in good agreement with quantum-mechanical computations, despite minor discrepancies in shape or intensity.
Demonstrating the extrapolative power of our best ML model, we studied the $\alpha$- to $\beta$-cristobalite transition, tracked via MAS spectra without the prohibitive computational cost of brute-force GIPAW calculations. Our analysis, leveraging thermal averages, revealed distinct NMR signatures between the two phases that are compatible with available experimental data.

Our ML framework makes it possible to estimate, from atomic coordinates alone, essential spectroscopic parameters in order to probe the local structure and the dynamics of (dis)ordered phases.  The accuracy of our ML models opens the door to systematically studying long-standing questions within the NMR community, such as the role of the antisymmetric component of the MS tensors in relaxation processes~\cite{anet_nmr_1990} and complex 2D NMR simulations to study dynamic processes, and subtle structural rearrangements~\cite{ashbrook_recent_2014,gm_rankin_advances_2023}.
While we focus on NMR properties in the present work, the formulation is more general, and we expect that the same ML approach can easily be extended to study other anisotropic (tensorial) material properties or electronic-structure observables.

We expect that, as we are using a GNN architecture, we can effortlessly extend the scope of our anisotropic NMR ML models to include additional chemical species without the need to retrain models from scratch. GNNs embed chemical species in their internal features, incorporating new chemical information does not result in an increase in the number of models' parameters, and hence the expansion comes with no additional execution performance cost. 
Moreover, combining our models with MLIPs accelerates the generation and the study of large structural models and the dynamic behavior of phase transitions in crystals. We view this as a key step toward more impactful modeling of materials, allowing for experimentally verifiable spectroscopic fingerprints in conjunction with increasingly accurate MLIPs (including recently reported models for more complex scenarios in the Si--O system \cite{erhard_modelling_2024,trillot_elaboration_2024,roy_learning_2024}). 

Looking further ahead, the use of flexible learning architectures, such as GNNs, paves the way to creating foundation models in a broad sense of the term -- capable of predicting several atomistic properties in the same inference step, as discussed in Ref.~\citenum{ben_mahmoud_data_2024}, and implemented for molecular systems~\cite{unke_physnet_2019}. 
An equally intriguing approach is to transfer learned GNN representations from one property to another -- for example, using the representations trained on the potential energy to enhance training efficiency for NMR parameters or electronic band gaps, and vice versa. Pilot studies for isotropic chemical shielding~\cite{ivkovic_transfer_2024} in carbon-based materials already showed promising results. Studying these correlations is important to maximize the leverage of atomistic foundation models and to extend their impact beyond their original target properties (typically, energies and forces), allowing for more comprehensive and more efficient materials discovery and design.

\section*{Data and code availability}
Data and code to reproduce the results will be made available in a public repository upon acceptance of this manuscript.

\begin{acknowledgments}
We acknowledge pilot studies by Jake Turner and Yixuan Huang at early stages of this work.
We thank Ilyes Batatia for early discussions and Daniel F.\ Thomas du Toit for help with XPOT usage (cf.~Ref.~\citenum{thomas_du_toit_cross-platform_2023}).
C.B.M. acknowledges funding from the Swiss National Science Foundation (SNSF) under grant number 217837.
This work was supported through a UK Research and Innovation Frontier Research grant [grant number EP/X016188/1].
This work was facilitated by software tools (specifically Soprano) developed by the Collaborative Computing Project for NMR Crystallography, funded by EPSRC grant EP/T026642/1.
We are grateful for computational support from the UK national high performance computing service, ARCHER2, for which access was obtained via the UKCP consortium and funded by EPSRC grant ref EP/X035891/1.
\end{acknowledgments}
\bibliographystyle{aipnum4-2}

\begin{thebibliography}{89}%
\makeatletter
\providecommand \@ifxundefined [1]{%
 \@ifx{#1\undefined}
}%
\providecommand \@ifnum [1]{%
 \ifnum #1\expandafter \@firstoftwo
 \else \expandafter \@secondoftwo
 \fi
}%
\providecommand \@ifx [1]{%
 \ifx #1\expandafter \@firstoftwo
 \else \expandafter \@secondoftwo
 \fi
}%
\providecommand \natexlab [1]{#1}%
\providecommand \enquote  [1]{``#1''}%
\providecommand \bibnamefont  [1]{#1}%
\providecommand \bibfnamefont [1]{#1}%
\providecommand \citenamefont [1]{#1}%
\providecommand \href@noop [0]{\@secondoftwo}%
\providecommand \href [0]{\begingroup \@sanitize@url \@href}%
\providecommand \@href[1]{\@@startlink{#1}\@@href}%
\providecommand \@@href[1]{\endgroup#1\@@endlink}%
\providecommand \@sanitize@url [0]{\catcode `\\12\catcode `\$12\catcode
  `\&12\catcode `\#12\catcode `\^12\catcode `\_12\catcode `\%12\relax}%
\providecommand \@@startlink[1]{}%
\providecommand \@@endlink[0]{}%
\providecommand \url  [0]{\begingroup\@sanitize@url \@url }%
\providecommand \@url [1]{\endgroup\@href {#1}{\urlprefix }}%
\providecommand \urlprefix  [0]{URL }%
\providecommand \Eprint [0]{\href }%
\providecommand \doibase [0]{https://doi.org/}%
\providecommand \selectlanguage [0]{\@gobble}%
\providecommand \bibinfo  [0]{\@secondoftwo}%
\providecommand \bibfield  [0]{\@secondoftwo}%
\providecommand \translation [1]{[#1]}%
\providecommand \BibitemOpen [0]{}%
\providecommand \bibitemStop [0]{}%
\providecommand \bibitemNoStop [0]{.\EOS\space}%
\providecommand \EOS [0]{\spacefactor3000\relax}%
\providecommand \BibitemShut  [1]{\csname bibitem#1\endcsname}%
\let\auto@bib@innerbib\@empty
\bibitem [{\citenamefont {Reif}\ \emph {et~al.}(2021)\citenamefont {Reif},
  \citenamefont {Ashbrook}, \citenamefont {Emsley},\ and\ \citenamefont
  {Hong}}]{reif_solid-state_2021}%
  \BibitemOpen
  \bibfield  {author} {\bibinfo {author} {\bibfnamefont {B.}~\bibnamefont
  {Reif}}, \bibinfo {author} {\bibfnamefont {S.~E.}\ \bibnamefont {Ashbrook}},
  \bibinfo {author} {\bibfnamefont {L.}~\bibnamefont {Emsley}},\ and\ \bibinfo
  {author} {\bibfnamefont {M.}~\bibnamefont {Hong}},\ }\href
  {https://doi.org/10.1038/s43586-020-00002-1} {\bibfield  {journal} {\bibinfo
  {journal} {Nature Reviews Methods Primers}\ }\textbf {\bibinfo {volume}
  {1}},\ \bibinfo {pages} {2} (\bibinfo {year} {2021})}\BibitemShut {NoStop}%
\bibitem [{\citenamefont {Pickard}\ and\ \citenamefont
  {Mauri}(2002)}]{pickard_first-principles_2002}%
  \BibitemOpen
  \bibfield  {author} {\bibinfo {author} {\bibfnamefont {C.~J.}\ \bibnamefont
  {Pickard}}\ and\ \bibinfo {author} {\bibfnamefont {F.}~\bibnamefont
  {Mauri}},\ }\href {https://doi.org/10.1103/PhysRevLett.88.086403} {\bibfield
  {journal} {\bibinfo  {journal} {Physical Review Letters}\ }\textbf {\bibinfo
  {volume} {88}},\ \bibinfo {pages} {086403} (\bibinfo {year}
  {2002})}\BibitemShut {NoStop}%
\bibitem [{\citenamefont {Holmes}\ \emph {et~al.}(2024)\citenamefont {Holmes},
  \citenamefont {Schönzart}, \citenamefont {Philips}, \citenamefont {Kimball},
  \citenamefont {Termos}, \citenamefont {Altenhof}, \citenamefont {Xu},
  \citenamefont {O'Keefe}, \citenamefont {Autschbach},\ and\ \citenamefont
  {Schurko}}]{holmes_structure_2024}%
  \BibitemOpen
  \bibfield  {author} {\bibinfo {author} {\bibfnamefont {S.~T.}\ \bibnamefont
  {Holmes}}, \bibinfo {author} {\bibfnamefont {J.}~\bibnamefont {Schönzart}},
  \bibinfo {author} {\bibfnamefont {A.~B.}\ \bibnamefont {Philips}}, \bibinfo
  {author} {\bibfnamefont {J.~J.}\ \bibnamefont {Kimball}}, \bibinfo {author}
  {\bibfnamefont {S.}~\bibnamefont {Termos}}, \bibinfo {author} {\bibfnamefont
  {A.~R.}\ \bibnamefont {Altenhof}}, \bibinfo {author} {\bibfnamefont
  {Y.}~\bibnamefont {Xu}}, \bibinfo {author} {\bibfnamefont {C.~A.}\
  \bibnamefont {O'Keefe}}, \bibinfo {author} {\bibfnamefont {J.}~\bibnamefont
  {Autschbach}},\ and\ \bibinfo {author} {\bibfnamefont {R.~W.}\ \bibnamefont
  {Schurko}},\ }\href {https://doi.org/10.1039/D3SC06026H} {\bibfield
  {journal} {\bibinfo  {journal} {Chemical Science}\ }\textbf {\bibinfo
  {volume} {15}},\ \bibinfo {pages} {2181} (\bibinfo {year}
  {2024})}\BibitemShut {NoStop}%
\bibitem [{\citenamefont {Dawson}\ \emph {et~al.}(2024)\citenamefont {Dawson},
  \citenamefont {Clayton}, \citenamefont {Marshall}, \citenamefont {Guillou},
  \citenamefont {Walton},\ and\ \citenamefont
  {Ashbrook}}]{dawson_site-directed_2024}%
  \BibitemOpen
  \bibfield  {author} {\bibinfo {author} {\bibfnamefont {D.~M.}\ \bibnamefont
  {Dawson}}, \bibinfo {author} {\bibfnamefont {J.~A.}\ \bibnamefont {Clayton}},
  \bibinfo {author} {\bibfnamefont {T.~H.~D.}\ \bibnamefont {Marshall}},
  \bibinfo {author} {\bibfnamefont {N.}~\bibnamefont {Guillou}}, \bibinfo
  {author} {\bibfnamefont {R.~I.}\ \bibnamefont {Walton}},\ and\ \bibinfo
  {author} {\bibfnamefont {S.~E.}\ \bibnamefont {Ashbrook}},\ }\href
  {https://doi.org/10.1039/D3SC06924A} {\bibfield  {journal} {\bibinfo
  {journal} {Chemical Science}\ }\textbf {\bibinfo {volume} {15}},\ \bibinfo
  {pages} {4374} (\bibinfo {year} {2024})}\BibitemShut {NoStop}%
\bibitem [{\citenamefont {Spearing}, \citenamefont {Farnan},\ and\
  \citenamefont {Stebbins}(1992)}]{spearing_dynamics_1992}%
  \BibitemOpen
  \bibfield  {author} {\bibinfo {author} {\bibfnamefont {D.}~\bibnamefont
  {Spearing}}, \bibinfo {author} {\bibfnamefont {I.}~\bibnamefont {Farnan}},\
  and\ \bibinfo {author} {\bibfnamefont {J.}~\bibnamefont {Stebbins}},\ }\href
  {https://doi.org/10.1007/BF00204008} {\bibfield  {journal} {\bibinfo
  {journal} {Physics and Chemistry of Minerals}\ }\textbf {\bibinfo {volume}
  {19}} (\bibinfo {year} {1992}),\ 10.1007/BF00204008}\BibitemShut {NoStop}%
\bibitem [{\citenamefont {Sattig}\ \emph {et~al.}(2014)\citenamefont {Sattig},
  \citenamefont {Reutter}, \citenamefont {Fujara}, \citenamefont {Werner},
  \citenamefont {Buntkowsky},\ and\ \citenamefont {Vogel}}]{sattig_nmr_2014}%
  \BibitemOpen
  \bibfield  {author} {\bibinfo {author} {\bibfnamefont {M.}~\bibnamefont
  {Sattig}}, \bibinfo {author} {\bibfnamefont {S.}~\bibnamefont {Reutter}},
  \bibinfo {author} {\bibfnamefont {F.}~\bibnamefont {Fujara}}, \bibinfo
  {author} {\bibfnamefont {M.}~\bibnamefont {Werner}}, \bibinfo {author}
  {\bibfnamefont {G.}~\bibnamefont {Buntkowsky}},\ and\ \bibinfo {author}
  {\bibfnamefont {M.}~\bibnamefont {Vogel}},\ }\href
  {https://doi.org/10.1039/C4CP02057J} {\bibfield  {journal} {\bibinfo
  {journal} {Phys. Chem. Chem. Phys.}\ }\textbf {\bibinfo {volume} {16}},\
  \bibinfo {pages} {19229} (\bibinfo {year} {2014})}\BibitemShut {NoStop}%
\bibitem [{\citenamefont {Huang}\ \emph {et~al.}(2022)\citenamefont {Huang},
  \citenamefont {Lee}, \citenamefont {Chang}, \citenamefont {Hong},
  \citenamefont {Ou}, \citenamefont {Kuo},\ and\ \citenamefont
  {Lue}}]{huang_27_2022}%
  \BibitemOpen
  \bibfield  {author} {\bibinfo {author} {\bibfnamefont {C.~Y.}\ \bibnamefont
  {Huang}}, \bibinfo {author} {\bibfnamefont {H.~Y.}\ \bibnamefont {Lee}},
  \bibinfo {author} {\bibfnamefont {Y.~C.}\ \bibnamefont {Chang}}, \bibinfo
  {author} {\bibfnamefont {C.~K.}\ \bibnamefont {Hong}}, \bibinfo {author}
  {\bibfnamefont {Y.~R.}\ \bibnamefont {Ou}}, \bibinfo {author} {\bibfnamefont
  {C.~N.}\ \bibnamefont {Kuo}},\ and\ \bibinfo {author} {\bibfnamefont {C.~S.}\
  \bibnamefont {Lue}},\ }\href {https://doi.org/10.1103/PhysRevB.106.195101}
  {\bibfield  {journal} {\bibinfo  {journal} {Physical Review B}\ }\textbf
  {\bibinfo {volume} {106}},\ \bibinfo {pages} {195101} (\bibinfo {year}
  {2022})}\BibitemShut {NoStop}%
\bibitem [{\citenamefont {Polenova}, \citenamefont {Gupta},\ and\ \citenamefont
  {Goldbourt}(2015)}]{polenova_magic_2015}%
  \BibitemOpen
  \bibfield  {author} {\bibinfo {author} {\bibfnamefont {T.}~\bibnamefont
  {Polenova}}, \bibinfo {author} {\bibfnamefont {R.}~\bibnamefont {Gupta}},\
  and\ \bibinfo {author} {\bibfnamefont {A.}~\bibnamefont {Goldbourt}},\ }\href
  {https://doi.org/10.1021/ac504288u} {\bibfield  {journal} {\bibinfo
  {journal} {Analytical Chemistry}\ }\textbf {\bibinfo {volume} {87}},\
  \bibinfo {pages} {5458} (\bibinfo {year} {2015})}\BibitemShut {NoStop}%
\bibitem [{\citenamefont {Ashbrook}\ and\ \citenamefont
  {McKay}(2016)}]{ashbrook_combining_2016}%
  \BibitemOpen
  \bibfield  {author} {\bibinfo {author} {\bibfnamefont {S.~E.}\ \bibnamefont
  {Ashbrook}}\ and\ \bibinfo {author} {\bibfnamefont {D.}~\bibnamefont
  {McKay}},\ }\href {https://doi.org/10.1039/C6CC02542K} {\bibfield  {journal}
  {\bibinfo  {journal} {Chemical Communications}\ }\textbf {\bibinfo {volume}
  {52}},\ \bibinfo {pages} {7186} (\bibinfo {year} {2016})}\BibitemShut
  {NoStop}%
\bibitem [{\citenamefont {Valenzuela~Reina}\ \emph {et~al.}(2024)\citenamefont
  {Valenzuela~Reina}, \citenamefont {Civaia}, \citenamefont {Harper},
  \citenamefont {Scheurer},\ and\ \citenamefont
  {Köcher}}]{valenzuela_reina_efg_2024}%
  \BibitemOpen
  \bibfield  {author} {\bibinfo {author} {\bibfnamefont {J.}~\bibnamefont
  {Valenzuela~Reina}}, \bibinfo {author} {\bibfnamefont {F.}~\bibnamefont
  {Civaia}}, \bibinfo {author} {\bibfnamefont {A.~F.}\ \bibnamefont {Harper}},
  \bibinfo {author} {\bibfnamefont {C.}~\bibnamefont {Scheurer}},\ and\
  \bibinfo {author} {\bibfnamefont {S.~S.}\ \bibnamefont {Köcher}},\ }\href
  {https://doi.org/10.1039/D4FD00075G} {\bibfield  {journal} {\bibinfo
  {journal} {Faraday Discussions}\ ,\ } (\bibinfo {year} {2024})},\ \bibinfo
  {note} {publisher: The Royal Society of Chemistry}\BibitemShut {NoStop}%
\bibitem [{\citenamefont {Harper}\ \emph {et~al.}(2023)\citenamefont {Harper},
  \citenamefont {Emge}, \citenamefont {Magusin}, \citenamefont {Grey},\ and\
  \citenamefont {Morris}}]{harper_modelling_2023}%
  \BibitemOpen
  \bibfield  {author} {\bibinfo {author} {\bibfnamefont {A.~F.}\ \bibnamefont
  {Harper}}, \bibinfo {author} {\bibfnamefont {S.~P.}\ \bibnamefont {Emge}},
  \bibinfo {author} {\bibfnamefont {P.~C. M.~M.}\ \bibnamefont {Magusin}},
  \bibinfo {author} {\bibfnamefont {C.~P.}\ \bibnamefont {Grey}},\ and\
  \bibinfo {author} {\bibfnamefont {A.~J.}\ \bibnamefont {Morris}},\ }\href
  {https://doi.org/10.1039/D2SC04035B} {\bibfield  {journal} {\bibinfo
  {journal} {Chemical Science}\ }\textbf {\bibinfo {volume} {14}},\ \bibinfo
  {pages} {1155} (\bibinfo {year} {2023})}\BibitemShut {NoStop}%
\bibitem [{\citenamefont {Pickard}\ and\ \citenamefont
  {Mauri}(2001)}]{pickard_all-electron_2001}%
  \BibitemOpen
  \bibfield  {author} {\bibinfo {author} {\bibfnamefont {C.~J.}\ \bibnamefont
  {Pickard}}\ and\ \bibinfo {author} {\bibfnamefont {F.}~\bibnamefont
  {Mauri}},\ }\href {https://doi.org/10.1103/PhysRevB.63.245101} {\bibfield
  {journal} {\bibinfo  {journal} {Physical Review B}\ }\textbf {\bibinfo
  {volume} {63}},\ \bibinfo {pages} {245101} (\bibinfo {year}
  {2001})}\BibitemShut {NoStop}%
\bibitem [{\citenamefont {Yates}, \citenamefont {Pickard},\ and\ \citenamefont
  {Mauri}(2007)}]{yates_calculation_2007}%
  \BibitemOpen
  \bibfield  {author} {\bibinfo {author} {\bibfnamefont {J.~R.}\ \bibnamefont
  {Yates}}, \bibinfo {author} {\bibfnamefont {C.~J.}\ \bibnamefont {Pickard}},\
  and\ \bibinfo {author} {\bibfnamefont {F.}~\bibnamefont {Mauri}},\ }\href
  {https://doi.org/10.1103/PhysRevB.76.024401} {\bibfield  {journal} {\bibinfo
  {journal} {Physical Review B}\ }\textbf {\bibinfo {volume} {76}},\ \bibinfo
  {pages} {024401} (\bibinfo {year} {2007})}\BibitemShut {NoStop}%
\bibitem [{\citenamefont {Behler}\ and\ \citenamefont
  {Parrinello}(2007)}]{behler_generalized_2007}%
  \BibitemOpen
  \bibfield  {author} {\bibinfo {author} {\bibfnamefont {J.}~\bibnamefont
  {Behler}}\ and\ \bibinfo {author} {\bibfnamefont {M.}~\bibnamefont
  {Parrinello}},\ }\href {https://doi.org/10/c7kbsq} {\bibfield  {journal}
  {\bibinfo  {journal} {Physical Review Letters}\ }\textbf {\bibinfo {volume}
  {98}},\ \bibinfo {pages} {146401} (\bibinfo {year} {2007})},\ \bibinfo {note}
  {publisher: American Physical Society}\BibitemShut {NoStop}%
\bibitem [{\citenamefont {Bartók}\ \emph {et~al.}(2010)\citenamefont
  {Bartók}, \citenamefont {Payne}, \citenamefont {Kondor},\ and\ \citenamefont
  {Csányi}}]{bartok_gaussian_2010}%
  \BibitemOpen
  \bibfield  {author} {\bibinfo {author} {\bibfnamefont {A.~P.}\ \bibnamefont
  {Bartók}}, \bibinfo {author} {\bibfnamefont {M.~C.}\ \bibnamefont {Payne}},
  \bibinfo {author} {\bibfnamefont {R.}~\bibnamefont {Kondor}},\ and\ \bibinfo
  {author} {\bibfnamefont {G.}~\bibnamefont {Csányi}},\ }\href
  {https://doi.org/10/dkchb7} {\bibfield  {journal} {\bibinfo  {journal}
  {Physical Review Letters}\ }\textbf {\bibinfo {volume} {104}},\ \bibinfo
  {pages} {136403} (\bibinfo {year} {2010})},\ \bibinfo {note} {publisher:
  American Physical Society}\BibitemShut {NoStop}%
\bibitem [{\citenamefont {Zhang}\ \emph {et~al.}(2018)\citenamefont {Zhang},
  \citenamefont {Han}, \citenamefont {Wang}, \citenamefont {Car},\ and\
  \citenamefont {E}}]{zhang_deep_2018}%
  \BibitemOpen
  \bibfield  {author} {\bibinfo {author} {\bibfnamefont {L.}~\bibnamefont
  {Zhang}}, \bibinfo {author} {\bibfnamefont {J.}~\bibnamefont {Han}}, \bibinfo
  {author} {\bibfnamefont {H.}~\bibnamefont {Wang}}, \bibinfo {author}
  {\bibfnamefont {R.}~\bibnamefont {Car}},\ and\ \bibinfo {author}
  {\bibfnamefont {W.}~\bibnamefont {E}},\ }\href
  {https://doi.org/10.1103/PhysRevLett.120.143001} {\bibfield  {journal}
  {\bibinfo  {journal} {Physical Review Letters}\ }\textbf {\bibinfo {volume}
  {120}},\ \bibinfo {pages} {143001} (\bibinfo {year} {2018})}\BibitemShut
  {NoStop}%
\bibitem [{\citenamefont {Drautz}(2019)}]{drautz_atomic_2019}%
  \BibitemOpen
  \bibfield  {author} {\bibinfo {author} {\bibfnamefont {R.}~\bibnamefont
  {Drautz}},\ }\href {https://doi.org/10.1103/PhysRevB.99.014104} {\bibfield
  {journal} {\bibinfo  {journal} {Physical Review B}\ }\textbf {\bibinfo
  {volume} {99}},\ \bibinfo {pages} {014104} (\bibinfo {year} {2019})},\
  \bibinfo {note} {publisher: American Physical Society}\BibitemShut {NoStop}%
\bibitem [{\citenamefont {Schütt}\ \emph {et~al.}(2019)\citenamefont
  {Schütt}, \citenamefont {Kessel}, \citenamefont {Gastegger}, \citenamefont
  {Nicoli}, \citenamefont {Tkatchenko},\ and\ \citenamefont
  {Müller}}]{schutt_schnetpack_2019}%
  \BibitemOpen
  \bibfield  {author} {\bibinfo {author} {\bibfnamefont {K.~T.}\ \bibnamefont
  {Schütt}}, \bibinfo {author} {\bibfnamefont {P.}~\bibnamefont {Kessel}},
  \bibinfo {author} {\bibfnamefont {M.}~\bibnamefont {Gastegger}}, \bibinfo
  {author} {\bibfnamefont {K.~A.}\ \bibnamefont {Nicoli}}, \bibinfo {author}
  {\bibfnamefont {A.}~\bibnamefont {Tkatchenko}},\ and\ \bibinfo {author}
  {\bibfnamefont {K.-R.}\ \bibnamefont {Müller}},\ }\href
  {https://doi.org/10.1021/acs.jctc.8b00908} {\bibfield  {journal} {\bibinfo
  {journal} {Journal of Chemical Theory and Computation}\ }\textbf {\bibinfo
  {volume} {15}},\ \bibinfo {pages} {448} (\bibinfo {year} {2019})}\BibitemShut
  {NoStop}%
\bibitem [{\citenamefont {Li}, \citenamefont {Kermode},\ and\ \citenamefont
  {De~Vita}(2015)}]{li_molecular_2015}%
  \BibitemOpen
  \bibfield  {author} {\bibinfo {author} {\bibfnamefont {Z.}~\bibnamefont
  {Li}}, \bibinfo {author} {\bibfnamefont {J.~R.}\ \bibnamefont {Kermode}},\
  and\ \bibinfo {author} {\bibfnamefont {A.}~\bibnamefont {De~Vita}},\ }\href
  {https://doi.org/10.1103/PhysRevLett.114.096405} {\bibfield  {journal}
  {\bibinfo  {journal} {Physical Review Letters}\ }\textbf {\bibinfo {volume}
  {114}},\ \bibinfo {pages} {096405} (\bibinfo {year} {2015})}\BibitemShut
  {NoStop}%
\bibitem [{\citenamefont {Unke}\ \emph {et~al.}(2021)\citenamefont {Unke},
  \citenamefont {Chmiela}, \citenamefont {Sauceda}, \citenamefont {Gastegger},
  \citenamefont {Poltavsky}, \citenamefont {Schütt}, \citenamefont
  {Tkatchenko},\ and\ \citenamefont {Müller}}]{unke_machine_2021}%
  \BibitemOpen
  \bibfield  {author} {\bibinfo {author} {\bibfnamefont {O.~T.}\ \bibnamefont
  {Unke}}, \bibinfo {author} {\bibfnamefont {S.}~\bibnamefont {Chmiela}},
  \bibinfo {author} {\bibfnamefont {H.~E.}\ \bibnamefont {Sauceda}}, \bibinfo
  {author} {\bibfnamefont {M.}~\bibnamefont {Gastegger}}, \bibinfo {author}
  {\bibfnamefont {I.}~\bibnamefont {Poltavsky}}, \bibinfo {author}
  {\bibfnamefont {K.~T.}\ \bibnamefont {Schütt}}, \bibinfo {author}
  {\bibfnamefont {A.}~\bibnamefont {Tkatchenko}},\ and\ \bibinfo {author}
  {\bibfnamefont {K.-R.}\ \bibnamefont {Müller}},\ }\href
  {https://doi.org/10.1021/acs.chemrev.0c01111} {\bibfield  {journal} {\bibinfo
   {journal} {Chemical Reviews}\ }\textbf {\bibinfo {volume} {121}},\ \bibinfo
  {pages} {10142} (\bibinfo {year} {2021})}\BibitemShut {NoStop}%
\bibitem [{\citenamefont {Deringer}, \citenamefont {Caro},\ and\ \citenamefont
  {Csányi}(2019)}]{deringer_machine_2019}%
  \BibitemOpen
  \bibfield  {author} {\bibinfo {author} {\bibfnamefont {V.~L.}\ \bibnamefont
  {Deringer}}, \bibinfo {author} {\bibfnamefont {M.~A.}\ \bibnamefont {Caro}},\
  and\ \bibinfo {author} {\bibfnamefont {G.}~\bibnamefont {Csányi}},\ }\href
  {https://doi.org/10.1002/adma.201902765} {\bibfield  {journal} {\bibinfo
  {journal} {Advanced Materials}\ }\textbf {\bibinfo {volume} {31}},\ \bibinfo
  {pages} {1902765} (\bibinfo {year} {2019})}\BibitemShut {NoStop}%
\bibitem [{\citenamefont {Cheng}\ \emph {et~al.}(2020)\citenamefont {Cheng},
  \citenamefont {Mazzola}, \citenamefont {Pickard},\ and\ \citenamefont
  {Ceriotti}}]{cheng_evidence_2020}%
  \BibitemOpen
  \bibfield  {author} {\bibinfo {author} {\bibfnamefont {B.}~\bibnamefont
  {Cheng}}, \bibinfo {author} {\bibfnamefont {G.}~\bibnamefont {Mazzola}},
  \bibinfo {author} {\bibfnamefont {C.~J.}\ \bibnamefont {Pickard}},\ and\
  \bibinfo {author} {\bibfnamefont {M.}~\bibnamefont {Ceriotti}},\ }\href
  {https://doi.org/10/gg99h2} {\bibfield  {journal} {\bibinfo  {journal}
  {Nature}\ }\textbf {\bibinfo {volume} {585}},\ \bibinfo {pages} {217}
  (\bibinfo {year} {2020})},\ \bibinfo {note} {number: 7824 Publisher: Nature
  Publishing Group}\BibitemShut {NoStop}%
\bibitem [{\citenamefont {Deringer}\ \emph {et~al.}(2021)\citenamefont
  {Deringer}, \citenamefont {Bernstein}, \citenamefont {Csányi}, \citenamefont
  {Ben~Mahmoud}, \citenamefont {Ceriotti}, \citenamefont {Wilson},
  \citenamefont {Drabold},\ and\ \citenamefont
  {Elliott}}]{deringer_origins_2021}%
  \BibitemOpen
  \bibfield  {author} {\bibinfo {author} {\bibfnamefont {V.~L.}\ \bibnamefont
  {Deringer}}, \bibinfo {author} {\bibfnamefont {N.}~\bibnamefont {Bernstein}},
  \bibinfo {author} {\bibfnamefont {G.}~\bibnamefont {Csányi}}, \bibinfo
  {author} {\bibfnamefont {C.}~\bibnamefont {Ben~Mahmoud}}, \bibinfo {author}
  {\bibfnamefont {M.}~\bibnamefont {Ceriotti}}, \bibinfo {author}
  {\bibfnamefont {M.}~\bibnamefont {Wilson}}, \bibinfo {author} {\bibfnamefont
  {D.~A.}\ \bibnamefont {Drabold}},\ and\ \bibinfo {author} {\bibfnamefont
  {S.~R.}\ \bibnamefont {Elliott}},\ }\href {https://doi.org/10/ghsb84}
  {\bibfield  {journal} {\bibinfo  {journal} {Nature}\ }\textbf {\bibinfo
  {volume} {589}},\ \bibinfo {pages} {59} (\bibinfo {year} {2021})},\ \bibinfo
  {note} {number: 7840 Publisher: Nature Publishing Group}\BibitemShut
  {NoStop}%
\bibitem [{\citenamefont {Deng}\ \emph
  {et~al.}(2023{\natexlab{a}})\citenamefont {Deng}, \citenamefont {Wang},
  \citenamefont {Xu},\ and\ \citenamefont {Li}}]{deng_machine_2023}%
  \BibitemOpen
  \bibfield  {author} {\bibinfo {author} {\bibfnamefont {Y.}~\bibnamefont
  {Deng}}, \bibinfo {author} {\bibfnamefont {C.}~\bibnamefont {Wang}}, \bibinfo
  {author} {\bibfnamefont {X.}~\bibnamefont {Xu}},\ and\ \bibinfo {author}
  {\bibfnamefont {H.}~\bibnamefont {Li}},\ }\href
  {https://doi.org/10.1016/j.taml.2023.100481} {\bibfield  {journal} {\bibinfo
  {journal} {Theoretical and Applied Mechanics Letters}\ }\textbf {\bibinfo
  {volume} {13}},\ \bibinfo {pages} {100481} (\bibinfo {year}
  {2023}{\natexlab{a}})}\BibitemShut {NoStop}%
\bibitem [{\citenamefont {Sosso}\ \emph {et~al.}(2013)\citenamefont {Sosso},
  \citenamefont {Miceli}, \citenamefont {Caravati}, \citenamefont {Giberti},
  \citenamefont {Behler},\ and\ \citenamefont {Bernasconi}}]{sosso_fast_2013}%
  \BibitemOpen
  \bibfield  {author} {\bibinfo {author} {\bibfnamefont {G.~C.}\ \bibnamefont
  {Sosso}}, \bibinfo {author} {\bibfnamefont {G.}~\bibnamefont {Miceli}},
  \bibinfo {author} {\bibfnamefont {S.}~\bibnamefont {Caravati}}, \bibinfo
  {author} {\bibfnamefont {F.}~\bibnamefont {Giberti}}, \bibinfo {author}
  {\bibfnamefont {J.}~\bibnamefont {Behler}},\ and\ \bibinfo {author}
  {\bibfnamefont {M.}~\bibnamefont {Bernasconi}},\ }\href
  {https://doi.org/10.1021/jz402268v} {\bibfield  {journal} {\bibinfo
  {journal} {The Journal of Physical Chemistry Letters}\ }\textbf {\bibinfo
  {volume} {4}},\ \bibinfo {pages} {4241} (\bibinfo {year} {2013})}\BibitemShut
  {NoStop}%
\bibitem [{\citenamefont {Piaggi}\ \emph {et~al.}(2022)\citenamefont {Piaggi},
  \citenamefont {Weis}, \citenamefont {Panagiotopoulos}, \citenamefont
  {Debenedetti},\ and\ \citenamefont {Car}}]{piaggi_homogeneous_2022}%
  \BibitemOpen
  \bibfield  {author} {\bibinfo {author} {\bibfnamefont {P.~M.}\ \bibnamefont
  {Piaggi}}, \bibinfo {author} {\bibfnamefont {J.}~\bibnamefont {Weis}},
  \bibinfo {author} {\bibfnamefont {A.~Z.}\ \bibnamefont {Panagiotopoulos}},
  \bibinfo {author} {\bibfnamefont {P.~G.}\ \bibnamefont {Debenedetti}},\ and\
  \bibinfo {author} {\bibfnamefont {R.}~\bibnamefont {Car}},\ }\href
  {https://doi.org/10.1073/pnas.2207294119} {\bibfield  {journal} {\bibinfo
  {journal} {Proceedings of the National Academy of Sciences}\ }\textbf
  {\bibinfo {volume} {119}},\ \bibinfo {pages} {e2207294119} (\bibinfo {year}
  {2022})}\BibitemShut {NoStop}%
\bibitem [{\citenamefont {Rupp}\ \emph {et~al.}(2012)\citenamefont {Rupp},
  \citenamefont {Tkatchenko}, \citenamefont {Müller},\ and\ \citenamefont {von
  Lilienfeld}}]{rupp_fast_2012}%
  \BibitemOpen
  \bibfield  {author} {\bibinfo {author} {\bibfnamefont {M.}~\bibnamefont
  {Rupp}}, \bibinfo {author} {\bibfnamefont {A.}~\bibnamefont {Tkatchenko}},
  \bibinfo {author} {\bibfnamefont {K.-R.}\ \bibnamefont {Müller}},\ and\
  \bibinfo {author} {\bibfnamefont {O.~A.}\ \bibnamefont {von Lilienfeld}},\
  }\href {https://doi.org/10.1103/PhysRevLett.108.058301} {\bibfield  {journal}
  {\bibinfo  {journal} {Physical Review Letters}\ }\textbf {\bibinfo {volume}
  {108}},\ \bibinfo {pages} {058301} (\bibinfo {year} {2012})}\BibitemShut
  {NoStop}%
\bibitem [{\citenamefont {Talapatra}\ \emph {et~al.}(2023)\citenamefont
  {Talapatra}, \citenamefont {Uberuaga}, \citenamefont {Stanek},\ and\
  \citenamefont {Pilania}}]{talapatra_band_2023}%
  \BibitemOpen
  \bibfield  {author} {\bibinfo {author} {\bibfnamefont {A.}~\bibnamefont
  {Talapatra}}, \bibinfo {author} {\bibfnamefont {B.~P.}\ \bibnamefont
  {Uberuaga}}, \bibinfo {author} {\bibfnamefont {C.~R.}\ \bibnamefont
  {Stanek}},\ and\ \bibinfo {author} {\bibfnamefont {G.}~\bibnamefont
  {Pilania}},\ }\href {https://doi.org/10.1038/s43246-023-00373-4} {\bibfield
  {journal} {\bibinfo  {journal} {Communications Materials}\ }\textbf {\bibinfo
  {volume} {4}},\ \bibinfo {pages} {46} (\bibinfo {year} {2023})}\BibitemShut
  {NoStop}%
\bibitem [{\citenamefont {Isayev}\ \emph {et~al.}(2017)\citenamefont {Isayev},
  \citenamefont {Oses}, \citenamefont {Toher}, \citenamefont {Gossett},
  \citenamefont {Curtarolo},\ and\ \citenamefont
  {Tropsha}}]{isayev_universal_2017}%
  \BibitemOpen
  \bibfield  {author} {\bibinfo {author} {\bibfnamefont {O.}~\bibnamefont
  {Isayev}}, \bibinfo {author} {\bibfnamefont {C.}~\bibnamefont {Oses}},
  \bibinfo {author} {\bibfnamefont {C.}~\bibnamefont {Toher}}, \bibinfo
  {author} {\bibfnamefont {E.}~\bibnamefont {Gossett}}, \bibinfo {author}
  {\bibfnamefont {S.}~\bibnamefont {Curtarolo}},\ and\ \bibinfo {author}
  {\bibfnamefont {A.}~\bibnamefont {Tropsha}},\ }\href
  {https://doi.org/10.1038/ncomms15679} {\bibfield  {journal} {\bibinfo
  {journal} {Nature Communications}\ }\textbf {\bibinfo {volume} {8}},\
  \bibinfo {pages} {15679} (\bibinfo {year} {2017})}\BibitemShut {NoStop}%
\bibitem [{\citenamefont {Wilkins}\ \emph {et~al.}(2019)\citenamefont
  {Wilkins}, \citenamefont {Grisafi}, \citenamefont {Yang}, \citenamefont
  {Lao}, \citenamefont {DiStasio},\ and\ \citenamefont
  {Ceriotti}}]{wilkins_accurate_2019}%
  \BibitemOpen
  \bibfield  {author} {\bibinfo {author} {\bibfnamefont {D.~M.}\ \bibnamefont
  {Wilkins}}, \bibinfo {author} {\bibfnamefont {A.}~\bibnamefont {Grisafi}},
  \bibinfo {author} {\bibfnamefont {Y.}~\bibnamefont {Yang}}, \bibinfo {author}
  {\bibfnamefont {K.~U.}\ \bibnamefont {Lao}}, \bibinfo {author} {\bibfnamefont
  {R.~A.}\ \bibnamefont {DiStasio}},\ and\ \bibinfo {author} {\bibfnamefont
  {M.}~\bibnamefont {Ceriotti}},\ }\href
  {https://doi.org/10.1073/pnas.1816132116} {\bibfield  {journal} {\bibinfo
  {journal} {Proceedings of the National Academy of Sciences}\ }\textbf
  {\bibinfo {volume} {116}},\ \bibinfo {pages} {3401} (\bibinfo {year}
  {2019})}\BibitemShut {NoStop}%
\bibitem [{\citenamefont {Grumet}\ \emph {et~al.}(2024)\citenamefont {Grumet},
  \citenamefont {Von~Scarpatetti}, \citenamefont {Bučko},\ and\ \citenamefont
  {Egger}}]{grumet_delta_2024}%
  \BibitemOpen
  \bibfield  {author} {\bibinfo {author} {\bibfnamefont {M.}~\bibnamefont
  {Grumet}}, \bibinfo {author} {\bibfnamefont {C.}~\bibnamefont
  {Von~Scarpatetti}}, \bibinfo {author} {\bibfnamefont {T.}~\bibnamefont
  {Bučko}},\ and\ \bibinfo {author} {\bibfnamefont {D.~A.}\ \bibnamefont
  {Egger}},\ }\href {https://doi.org/10.1021/acs.jpcc.4c00886} {\bibfield
  {journal} {\bibinfo  {journal} {The Journal of Physical Chemistry C}\
  }\textbf {\bibinfo {volume} {128}},\ \bibinfo {pages} {6464} (\bibinfo {year}
  {2024})}\BibitemShut {NoStop}%
\bibitem [{\citenamefont {Gigli}\ \emph {et~al.}(2022)\citenamefont {Gigli},
  \citenamefont {Veit}, \citenamefont {Kotiuga}, \citenamefont {Pizzi},
  \citenamefont {Marzari},\ and\ \citenamefont
  {Ceriotti}}]{gigli_thermodynamics_2022}%
  \BibitemOpen
  \bibfield  {author} {\bibinfo {author} {\bibfnamefont {L.}~\bibnamefont
  {Gigli}}, \bibinfo {author} {\bibfnamefont {M.}~\bibnamefont {Veit}},
  \bibinfo {author} {\bibfnamefont {M.}~\bibnamefont {Kotiuga}}, \bibinfo
  {author} {\bibfnamefont {G.}~\bibnamefont {Pizzi}}, \bibinfo {author}
  {\bibfnamefont {N.}~\bibnamefont {Marzari}},\ and\ \bibinfo {author}
  {\bibfnamefont {M.}~\bibnamefont {Ceriotti}},\ }\href
  {https://doi.org/10.1038/s41524-022-00845-0} {\bibfield  {journal} {\bibinfo
  {journal} {npj Computational Materials}\ }\textbf {\bibinfo {volume} {8}},\
  \bibinfo {pages} {209} (\bibinfo {year} {2022})}\BibitemShut {NoStop}%
\bibitem [{\citenamefont {Grisafi}\ \emph {et~al.}(2019)\citenamefont
  {Grisafi}, \citenamefont {Fabrizio}, \citenamefont {Meyer}, \citenamefont
  {Wilkins}, \citenamefont {Corminboeuf},\ and\ \citenamefont
  {Ceriotti}}]{grisafi_transferable_2019}%
  \BibitemOpen
  \bibfield  {author} {\bibinfo {author} {\bibfnamefont {A.}~\bibnamefont
  {Grisafi}}, \bibinfo {author} {\bibfnamefont {A.}~\bibnamefont {Fabrizio}},
  \bibinfo {author} {\bibfnamefont {B.}~\bibnamefont {Meyer}}, \bibinfo
  {author} {\bibfnamefont {D.~M.}\ \bibnamefont {Wilkins}}, \bibinfo {author}
  {\bibfnamefont {C.}~\bibnamefont {Corminboeuf}},\ and\ \bibinfo {author}
  {\bibfnamefont {M.}~\bibnamefont {Ceriotti}},\ }\href
  {https://doi.org/10.1021/acscentsci.8b00551} {\bibfield  {journal} {\bibinfo
  {journal} {ACS Central Science}\ }\textbf {\bibinfo {volume} {5}},\ \bibinfo
  {pages} {57} (\bibinfo {year} {2019})}\BibitemShut {NoStop}%
\bibitem [{\citenamefont {Chandrasekaran}\ \emph {et~al.}(2019)\citenamefont
  {Chandrasekaran}, \citenamefont {Kamal}, \citenamefont {Batra}, \citenamefont
  {Kim}, \citenamefont {Chen},\ and\ \citenamefont
  {Ramprasad}}]{chandrasekaran_solving_2019}%
  \BibitemOpen
  \bibfield  {author} {\bibinfo {author} {\bibfnamefont {A.}~\bibnamefont
  {Chandrasekaran}}, \bibinfo {author} {\bibfnamefont {D.}~\bibnamefont
  {Kamal}}, \bibinfo {author} {\bibfnamefont {R.}~\bibnamefont {Batra}},
  \bibinfo {author} {\bibfnamefont {C.}~\bibnamefont {Kim}}, \bibinfo {author}
  {\bibfnamefont {L.}~\bibnamefont {Chen}},\ and\ \bibinfo {author}
  {\bibfnamefont {R.}~\bibnamefont {Ramprasad}},\ }\href
  {https://doi.org/10.1038/s41524-019-0162-7} {\bibfield  {journal} {\bibinfo
  {journal} {npj Computational Materials}\ }\textbf {\bibinfo {volume} {5}},\
  \bibinfo {pages} {22} (\bibinfo {year} {2019})}\BibitemShut {NoStop}%
\bibitem [{\citenamefont {Ben~Mahmoud}\ \emph {et~al.}(2020)\citenamefont
  {Ben~Mahmoud}, \citenamefont {Anelli}, \citenamefont {Csányi},\ and\
  \citenamefont {Ceriotti}}]{ben_mahmoud_learning_2020}%
  \BibitemOpen
  \bibfield  {author} {\bibinfo {author} {\bibfnamefont {C.}~\bibnamefont
  {Ben~Mahmoud}}, \bibinfo {author} {\bibfnamefont {A.}~\bibnamefont {Anelli}},
  \bibinfo {author} {\bibfnamefont {G.}~\bibnamefont {Csányi}},\ and\ \bibinfo
  {author} {\bibfnamefont {M.}~\bibnamefont {Ceriotti}},\ }\href
  {https://doi.org/10.1103/PhysRevB.102.235130} {\bibfield  {journal} {\bibinfo
   {journal} {Physical Review B}\ }\textbf {\bibinfo {volume} {102}},\ \bibinfo
  {pages} {235130} (\bibinfo {year} {2020})},\ \bibinfo {note} {publisher:
  American Physical Society}\BibitemShut {NoStop}%
\bibitem [{\citenamefont {Dey}\ and\ \citenamefont
  {Ghosh}(2023)}]{dey_machine_2023}%
  \BibitemOpen
  \bibfield  {author} {\bibinfo {author} {\bibfnamefont {M.}~\bibnamefont
  {Dey}}\ and\ \bibinfo {author} {\bibfnamefont {D.}~\bibnamefont {Ghosh}},\
  }\href {https://doi.org/10.1021/acs.jpca.3c05322} {\bibfield  {journal}
  {\bibinfo  {journal} {The Journal of Physical Chemistry A}\ }\textbf
  {\bibinfo {volume} {127}},\ \bibinfo {pages} {9159} (\bibinfo {year}
  {2023})}\BibitemShut {NoStop}%
\bibitem [{\citenamefont {Fiedler}\ \emph {et~al.}(2023)\citenamefont
  {Fiedler}, \citenamefont {Modine}, \citenamefont {Schmerler}, \citenamefont
  {Vogel}, \citenamefont {Popoola}, \citenamefont {Thompson}, \citenamefont
  {Rajamanickam},\ and\ \citenamefont {Cangi}}]{fiedler_predicting_2023}%
  \BibitemOpen
  \bibfield  {author} {\bibinfo {author} {\bibfnamefont {L.}~\bibnamefont
  {Fiedler}}, \bibinfo {author} {\bibfnamefont {N.~A.}\ \bibnamefont {Modine}},
  \bibinfo {author} {\bibfnamefont {S.}~\bibnamefont {Schmerler}}, \bibinfo
  {author} {\bibfnamefont {D.~J.}\ \bibnamefont {Vogel}}, \bibinfo {author}
  {\bibfnamefont {G.~A.}\ \bibnamefont {Popoola}}, \bibinfo {author}
  {\bibfnamefont {A.~P.}\ \bibnamefont {Thompson}}, \bibinfo {author}
  {\bibfnamefont {S.}~\bibnamefont {Rajamanickam}},\ and\ \bibinfo {author}
  {\bibfnamefont {A.}~\bibnamefont {Cangi}},\ }\href
  {https://doi.org/10.1038/s41524-023-01070-z} {\bibfield  {journal} {\bibinfo
  {journal} {npj Computational Materials}\ }\textbf {\bibinfo {volume} {9}},\
  \bibinfo {pages} {115} (\bibinfo {year} {2023})}\BibitemShut {NoStop}%
\bibitem [{\citenamefont {Scherbela}, \citenamefont {Gerard},\ and\
  \citenamefont {Grohs}(2024)}]{scherbela_towards_2024}%
  \BibitemOpen
  \bibfield  {author} {\bibinfo {author} {\bibfnamefont {M.}~\bibnamefont
  {Scherbela}}, \bibinfo {author} {\bibfnamefont {L.}~\bibnamefont {Gerard}},\
  and\ \bibinfo {author} {\bibfnamefont {P.}~\bibnamefont {Grohs}},\ }\href
  {https://doi.org/10.1038/s41467-023-44216-9} {\bibfield  {journal} {\bibinfo
  {journal} {Nature Communications}\ }\textbf {\bibinfo {volume} {15}},\
  \bibinfo {pages} {120} (\bibinfo {year} {2024})}\BibitemShut {NoStop}%
\bibitem [{\citenamefont {Sun}\ \emph {et~al.}(2022)\citenamefont {Sun},
  \citenamefont {Xiang}, \citenamefont {Liu}, \citenamefont {Xu}, \citenamefont
  {Leng}, \citenamefont {Ye}, \citenamefont {Fortunelli}, \citenamefont
  {Goddard},\ and\ \citenamefont {Cheng}}]{sun_machine_2022}%
  \BibitemOpen
  \bibfield  {author} {\bibinfo {author} {\bibfnamefont {Q.}~\bibnamefont
  {Sun}}, \bibinfo {author} {\bibfnamefont {Y.}~\bibnamefont {Xiang}}, \bibinfo
  {author} {\bibfnamefont {Y.}~\bibnamefont {Liu}}, \bibinfo {author}
  {\bibfnamefont {L.}~\bibnamefont {Xu}}, \bibinfo {author} {\bibfnamefont
  {T.}~\bibnamefont {Leng}}, \bibinfo {author} {\bibfnamefont {Y.}~\bibnamefont
  {Ye}}, \bibinfo {author} {\bibfnamefont {A.}~\bibnamefont {Fortunelli}},
  \bibinfo {author} {\bibfnamefont {W.~A.}\ \bibnamefont {Goddard}},\ and\
  \bibinfo {author} {\bibfnamefont {T.}~\bibnamefont {Cheng}},\ }\href
  {https://doi.org/10.1021/acs.jpclett.2c02222} {\bibfield  {journal} {\bibinfo
   {journal} {The Journal of Physical Chemistry Letters}\ }\textbf {\bibinfo
  {volume} {13}},\ \bibinfo {pages} {8047} (\bibinfo {year}
  {2022})}\BibitemShut {NoStop}%
\bibitem [{\citenamefont {Golze}\ \emph {et~al.}(2022)\citenamefont {Golze},
  \citenamefont {Hirvensalo}, \citenamefont {Hernández-León}, \citenamefont
  {Aarva}, \citenamefont {Etula}, \citenamefont {Susi}, \citenamefont {Rinke},
  \citenamefont {Laurila},\ and\ \citenamefont {Caro}}]{golze_accurate_2022}%
  \BibitemOpen
  \bibfield  {author} {\bibinfo {author} {\bibfnamefont {D.}~\bibnamefont
  {Golze}}, \bibinfo {author} {\bibfnamefont {M.}~\bibnamefont {Hirvensalo}},
  \bibinfo {author} {\bibfnamefont {P.}~\bibnamefont {Hernández-León}},
  \bibinfo {author} {\bibfnamefont {A.}~\bibnamefont {Aarva}}, \bibinfo
  {author} {\bibfnamefont {J.}~\bibnamefont {Etula}}, \bibinfo {author}
  {\bibfnamefont {T.}~\bibnamefont {Susi}}, \bibinfo {author} {\bibfnamefont
  {P.}~\bibnamefont {Rinke}}, \bibinfo {author} {\bibfnamefont
  {T.}~\bibnamefont {Laurila}},\ and\ \bibinfo {author} {\bibfnamefont {M.~A.}\
  \bibnamefont {Caro}},\ }\href {https://doi.org/10.1021/acs.chemmater.1c04279}
  {\bibfield  {journal} {\bibinfo  {journal} {Chemistry of Materials}\ }\textbf
  {\bibinfo {volume} {34}},\ \bibinfo {pages} {6240} (\bibinfo {year}
  {2022})}\BibitemShut {NoStop}%
\bibitem [{\citenamefont {Batatia}\ \emph {et~al.}(2023)\citenamefont
  {Batatia}, \citenamefont {Benner}, \citenamefont {Chiang}, \citenamefont
  {Elena}, \citenamefont {Kovács}, \citenamefont {Riebesell}, \citenamefont
  {Advincula}, \citenamefont {Asta}, \citenamefont {Baldwin}, \citenamefont
  {Bernstein}, \citenamefont {Bhowmik}, \citenamefont {Blau}, \citenamefont
  {Cărare}, \citenamefont {Darby}, \citenamefont {De}, \citenamefont
  {Della~Pia}, \citenamefont {Deringer}, \citenamefont {Elijošius},
  \citenamefont {El-Machachi}, \citenamefont {Fako}, \citenamefont {Ferrari},
  \citenamefont {Genreith-Schriever}, \citenamefont {George}, \citenamefont
  {Goodall}, \citenamefont {Grey}, \citenamefont {Han}, \citenamefont
  {Handley}, \citenamefont {Heenen}, \citenamefont {Hermansson}, \citenamefont
  {Holm}, \citenamefont {Jaafar}, \citenamefont {Hofmann}, \citenamefont
  {Jakob}, \citenamefont {Jung}, \citenamefont {Kapil}, \citenamefont {Kaplan},
  \citenamefont {Karimitari}, \citenamefont {Kroupa}, \citenamefont {Kullgren},
  \citenamefont {Kuner}, \citenamefont {Kuryla}, \citenamefont {Liepuoniute},
  \citenamefont {Margraf}, \citenamefont {Magdău}, \citenamefont
  {Michaelides}, \citenamefont {Moore}, \citenamefont {Naik}, \citenamefont
  {Niblett}, \citenamefont {Norwood}, \citenamefont {O'Neill}, \citenamefont
  {Ortner}, \citenamefont {Persson}, \citenamefont {Reuter}, \citenamefont
  {Rosen}, \citenamefont {Schaaf}, \citenamefont {Schran}, \citenamefont
  {Sivonxay}, \citenamefont {Stenczel}, \citenamefont {Svahn}, \citenamefont
  {Sutton}, \citenamefont {van~der Oord}, \citenamefont {Varga-Umbrich},
  \citenamefont {Vegge}, \citenamefont {Vondrák}, \citenamefont {Wang},
  \citenamefont {Witt}, \citenamefont {Zills},\ and\ \citenamefont
  {Csányi}}]{batatia_foundation_2023}%
  \BibitemOpen
  \bibfield  {author} {\bibinfo {author} {\bibfnamefont {I.}~\bibnamefont
  {Batatia}}, \bibinfo {author} {\bibfnamefont {P.}~\bibnamefont {Benner}},
  \bibinfo {author} {\bibfnamefont {Y.}~\bibnamefont {Chiang}}, \bibinfo
  {author} {\bibfnamefont {A.~M.}\ \bibnamefont {Elena}}, \bibinfo {author}
  {\bibfnamefont {D.~P.}\ \bibnamefont {Kovács}}, \bibinfo {author}
  {\bibfnamefont {J.}~\bibnamefont {Riebesell}}, \bibinfo {author}
  {\bibfnamefont {X.~R.}\ \bibnamefont {Advincula}}, \bibinfo {author}
  {\bibfnamefont {M.}~\bibnamefont {Asta}}, \bibinfo {author} {\bibfnamefont
  {W.~J.}\ \bibnamefont {Baldwin}}, \bibinfo {author} {\bibfnamefont
  {N.}~\bibnamefont {Bernstein}}, \bibinfo {author} {\bibfnamefont
  {A.}~\bibnamefont {Bhowmik}}, \bibinfo {author} {\bibfnamefont {S.~M.}\
  \bibnamefont {Blau}}, \bibinfo {author} {\bibfnamefont {V.}~\bibnamefont
  {Cărare}}, \bibinfo {author} {\bibfnamefont {J.~P.}\ \bibnamefont {Darby}},
  \bibinfo {author} {\bibfnamefont {S.}~\bibnamefont {De}}, \bibinfo {author}
  {\bibfnamefont {F.}~\bibnamefont {Della~Pia}}, \bibinfo {author}
  {\bibfnamefont {V.~L.}\ \bibnamefont {Deringer}}, \bibinfo {author}
  {\bibfnamefont {R.}~\bibnamefont {Elijošius}}, \bibinfo {author}
  {\bibfnamefont {Z.}~\bibnamefont {El-Machachi}}, \bibinfo {author}
  {\bibfnamefont {E.}~\bibnamefont {Fako}}, \bibinfo {author} {\bibfnamefont
  {A.~C.}\ \bibnamefont {Ferrari}}, \bibinfo {author} {\bibfnamefont
  {A.}~\bibnamefont {Genreith-Schriever}}, \bibinfo {author} {\bibfnamefont
  {J.}~\bibnamefont {George}}, \bibinfo {author} {\bibfnamefont {R.~E.~A.}\
  \bibnamefont {Goodall}}, \bibinfo {author} {\bibfnamefont {C.~P.}\
  \bibnamefont {Grey}}, \bibinfo {author} {\bibfnamefont {S.}~\bibnamefont
  {Han}}, \bibinfo {author} {\bibfnamefont {W.}~\bibnamefont {Handley}},
  \bibinfo {author} {\bibfnamefont {H.~H.}\ \bibnamefont {Heenen}}, \bibinfo
  {author} {\bibfnamefont {K.}~\bibnamefont {Hermansson}}, \bibinfo {author}
  {\bibfnamefont {C.}~\bibnamefont {Holm}}, \bibinfo {author} {\bibfnamefont
  {J.}~\bibnamefont {Jaafar}}, \bibinfo {author} {\bibfnamefont
  {S.}~\bibnamefont {Hofmann}}, \bibinfo {author} {\bibfnamefont {K.~S.}\
  \bibnamefont {Jakob}}, \bibinfo {author} {\bibfnamefont {H.}~\bibnamefont
  {Jung}}, \bibinfo {author} {\bibfnamefont {V.}~\bibnamefont {Kapil}},
  \bibinfo {author} {\bibfnamefont {A.~D.}\ \bibnamefont {Kaplan}}, \bibinfo
  {author} {\bibfnamefont {N.}~\bibnamefont {Karimitari}}, \bibinfo {author}
  {\bibfnamefont {N.}~\bibnamefont {Kroupa}}, \bibinfo {author} {\bibfnamefont
  {J.}~\bibnamefont {Kullgren}}, \bibinfo {author} {\bibfnamefont {M.~C.}\
  \bibnamefont {Kuner}}, \bibinfo {author} {\bibfnamefont {D.}~\bibnamefont
  {Kuryla}}, \bibinfo {author} {\bibfnamefont {G.}~\bibnamefont {Liepuoniute}},
  \bibinfo {author} {\bibfnamefont {J.~T.}\ \bibnamefont {Margraf}}, \bibinfo
  {author} {\bibfnamefont {I.-B.}\ \bibnamefont {Magdău}}, \bibinfo {author}
  {\bibfnamefont {A.}~\bibnamefont {Michaelides}}, \bibinfo {author}
  {\bibfnamefont {J.~H.}\ \bibnamefont {Moore}}, \bibinfo {author}
  {\bibfnamefont {A.~A.}\ \bibnamefont {Naik}}, \bibinfo {author}
  {\bibfnamefont {S.~P.}\ \bibnamefont {Niblett}}, \bibinfo {author}
  {\bibfnamefont {S.~W.}\ \bibnamefont {Norwood}}, \bibinfo {author}
  {\bibfnamefont {N.}~\bibnamefont {O'Neill}}, \bibinfo {author} {\bibfnamefont
  {C.}~\bibnamefont {Ortner}}, \bibinfo {author} {\bibfnamefont {K.~A.}\
  \bibnamefont {Persson}}, \bibinfo {author} {\bibfnamefont {K.}~\bibnamefont
  {Reuter}}, \bibinfo {author} {\bibfnamefont {A.~S.}\ \bibnamefont {Rosen}},
  \bibinfo {author} {\bibfnamefont {L.~L.}\ \bibnamefont {Schaaf}}, \bibinfo
  {author} {\bibfnamefont {C.}~\bibnamefont {Schran}}, \bibinfo {author}
  {\bibfnamefont {E.}~\bibnamefont {Sivonxay}}, \bibinfo {author}
  {\bibfnamefont {T.~K.}\ \bibnamefont {Stenczel}}, \bibinfo {author}
  {\bibfnamefont {V.}~\bibnamefont {Svahn}}, \bibinfo {author} {\bibfnamefont
  {C.}~\bibnamefont {Sutton}}, \bibinfo {author} {\bibfnamefont
  {C.}~\bibnamefont {van~der Oord}}, \bibinfo {author} {\bibfnamefont
  {E.}~\bibnamefont {Varga-Umbrich}}, \bibinfo {author} {\bibfnamefont
  {T.}~\bibnamefont {Vegge}}, \bibinfo {author} {\bibfnamefont
  {M.}~\bibnamefont {Vondrák}}, \bibinfo {author} {\bibfnamefont
  {Y.}~\bibnamefont {Wang}}, \bibinfo {author} {\bibfnamefont {W.~C.}\
  \bibnamefont {Witt}}, \bibinfo {author} {\bibfnamefont {F.}~\bibnamefont
  {Zills}},\ and\ \bibinfo {author} {\bibfnamefont {G.}~\bibnamefont
  {Csányi}},\ }\href {https://doi.org/10.48550/arXiv.2401.00096} {\enquote
  {\bibinfo {title} {A foundation model for atomistic materials chemistry},}\ }
  (\bibinfo {year} {2023}),\ \bibinfo {note} {arXiv:2401.00096 [cond-mat,
  physics:physics]}\BibitemShut {NoStop}%
\bibitem [{\citenamefont {Deng}\ \emph
  {et~al.}(2023{\natexlab{b}})\citenamefont {Deng}, \citenamefont {Zhong},
  \citenamefont {Jun}, \citenamefont {Riebesell}, \citenamefont {Han},
  \citenamefont {Bartel},\ and\ \citenamefont {Ceder}}]{deng_chgnet_2023}%
  \BibitemOpen
  \bibfield  {author} {\bibinfo {author} {\bibfnamefont {B.}~\bibnamefont
  {Deng}}, \bibinfo {author} {\bibfnamefont {P.}~\bibnamefont {Zhong}},
  \bibinfo {author} {\bibfnamefont {K.}~\bibnamefont {Jun}}, \bibinfo {author}
  {\bibfnamefont {J.}~\bibnamefont {Riebesell}}, \bibinfo {author}
  {\bibfnamefont {K.}~\bibnamefont {Han}}, \bibinfo {author} {\bibfnamefont
  {C.~J.}\ \bibnamefont {Bartel}},\ and\ \bibinfo {author} {\bibfnamefont
  {G.}~\bibnamefont {Ceder}},\ }\href
  {https://doi.org/10.1038/s42256-023-00716-3} {\bibfield  {journal} {\bibinfo
  {journal} {Nature Machine Intelligence}\ }\textbf {\bibinfo {volume} {5}},\
  \bibinfo {pages} {1031} (\bibinfo {year} {2023}{\natexlab{b}})}\BibitemShut
  {NoStop}%
\bibitem [{\citenamefont {Merchant}\ \emph {et~al.}(2023)\citenamefont
  {Merchant}, \citenamefont {Batzner}, \citenamefont {Schoenholz},
  \citenamefont {Aykol}, \citenamefont {Cheon},\ and\ \citenamefont
  {Cubuk}}]{merchant_scaling_2023}%
  \BibitemOpen
  \bibfield  {author} {\bibinfo {author} {\bibfnamefont {A.}~\bibnamefont
  {Merchant}}, \bibinfo {author} {\bibfnamefont {S.}~\bibnamefont {Batzner}},
  \bibinfo {author} {\bibfnamefont {S.~S.}\ \bibnamefont {Schoenholz}},
  \bibinfo {author} {\bibfnamefont {M.}~\bibnamefont {Aykol}}, \bibinfo
  {author} {\bibfnamefont {G.}~\bibnamefont {Cheon}},\ and\ \bibinfo {author}
  {\bibfnamefont {E.~D.}\ \bibnamefont {Cubuk}},\ }\href
  {https://doi.org/10.1038/s41586-023-06735-9} {\bibfield  {journal} {\bibinfo
  {journal} {Nature}\ }\textbf {\bibinfo {volume} {624}},\ \bibinfo {pages}
  {80} (\bibinfo {year} {2023})}\BibitemShut {NoStop}%
\bibitem [{\citenamefont {Yang}\ \emph {et~al.}(2024)\citenamefont {Yang},
  \citenamefont {Hu}, \citenamefont {Zhou}, \citenamefont {Liu}, \citenamefont
  {Shi}, \citenamefont {Li}, \citenamefont {Li}, \citenamefont {Chen},
  \citenamefont {Chen}, \citenamefont {Zeni}, \citenamefont {Horton},
  \citenamefont {Pinsler}, \citenamefont {Fowler}, \citenamefont {Zügner},
  \citenamefont {Xie}, \citenamefont {Smith}, \citenamefont {Sun},
  \citenamefont {Wang}, \citenamefont {Kong}, \citenamefont {Liu},
  \citenamefont {Hao},\ and\ \citenamefont {Lu}}]{yang_mattersim_2024}%
  \BibitemOpen
  \bibfield  {author} {\bibinfo {author} {\bibfnamefont {H.}~\bibnamefont
  {Yang}}, \bibinfo {author} {\bibfnamefont {C.}~\bibnamefont {Hu}}, \bibinfo
  {author} {\bibfnamefont {Y.}~\bibnamefont {Zhou}}, \bibinfo {author}
  {\bibfnamefont {X.}~\bibnamefont {Liu}}, \bibinfo {author} {\bibfnamefont
  {Y.}~\bibnamefont {Shi}}, \bibinfo {author} {\bibfnamefont {J.}~\bibnamefont
  {Li}}, \bibinfo {author} {\bibfnamefont {G.}~\bibnamefont {Li}}, \bibinfo
  {author} {\bibfnamefont {Z.}~\bibnamefont {Chen}}, \bibinfo {author}
  {\bibfnamefont {S.}~\bibnamefont {Chen}}, \bibinfo {author} {\bibfnamefont
  {C.}~\bibnamefont {Zeni}}, \bibinfo {author} {\bibfnamefont {M.}~\bibnamefont
  {Horton}}, \bibinfo {author} {\bibfnamefont {R.}~\bibnamefont {Pinsler}},
  \bibinfo {author} {\bibfnamefont {A.}~\bibnamefont {Fowler}}, \bibinfo
  {author} {\bibfnamefont {D.}~\bibnamefont {Zügner}}, \bibinfo {author}
  {\bibfnamefont {T.}~\bibnamefont {Xie}}, \bibinfo {author} {\bibfnamefont
  {J.}~\bibnamefont {Smith}}, \bibinfo {author} {\bibfnamefont
  {L.}~\bibnamefont {Sun}}, \bibinfo {author} {\bibfnamefont {Q.}~\bibnamefont
  {Wang}}, \bibinfo {author} {\bibfnamefont {L.}~\bibnamefont {Kong}}, \bibinfo
  {author} {\bibfnamefont {C.}~\bibnamefont {Liu}}, \bibinfo {author}
  {\bibfnamefont {H.}~\bibnamefont {Hao}},\ and\ \bibinfo {author}
  {\bibfnamefont {Z.}~\bibnamefont {Lu}},\ }\href
  {https://doi.org/10.48550/ARXIV.2405.04967} {\enquote {\bibinfo {title}
  {{MatterSim}: {A} {Deep} {Learning} {Atomistic} {Model} {Across} {Elements},
  {Temperatures} and {Pressures}},}\ } (\bibinfo {year} {2024}),\ \bibinfo
  {note} {version Number: 2}\BibitemShut {NoStop}%
\bibitem [{\citenamefont {Kwon}\ \emph {et~al.}(2020)\citenamefont {Kwon},
  \citenamefont {Lee}, \citenamefont {Choi}, \citenamefont {Kang},\ and\
  \citenamefont {Kang}}]{kwon_neural_2020}%
  \BibitemOpen
  \bibfield  {author} {\bibinfo {author} {\bibfnamefont {Y.}~\bibnamefont
  {Kwon}}, \bibinfo {author} {\bibfnamefont {D.}~\bibnamefont {Lee}}, \bibinfo
  {author} {\bibfnamefont {Y.-S.}\ \bibnamefont {Choi}}, \bibinfo {author}
  {\bibfnamefont {M.}~\bibnamefont {Kang}},\ and\ \bibinfo {author}
  {\bibfnamefont {S.}~\bibnamefont {Kang}},\ }\href
  {https://doi.org/10.1021/acs.jcim.0c00195} {\bibfield  {journal} {\bibinfo
  {journal} {Journal of Chemical Information and Modeling}\ }\textbf {\bibinfo
  {volume} {60}},\ \bibinfo {pages} {2024} (\bibinfo {year}
  {2020})}\BibitemShut {NoStop}%
\bibitem [{\citenamefont {Han}\ \emph {et~al.}(2022)\citenamefont {Han},
  \citenamefont {Kang}, \citenamefont {Kang}, \citenamefont {Kwon},
  \citenamefont {Lee},\ and\ \citenamefont {Choi}}]{han_scalable_2022}%
  \BibitemOpen
  \bibfield  {author} {\bibinfo {author} {\bibfnamefont {J.}~\bibnamefont
  {Han}}, \bibinfo {author} {\bibfnamefont {H.}~\bibnamefont {Kang}}, \bibinfo
  {author} {\bibfnamefont {S.}~\bibnamefont {Kang}}, \bibinfo {author}
  {\bibfnamefont {Y.}~\bibnamefont {Kwon}}, \bibinfo {author} {\bibfnamefont
  {D.}~\bibnamefont {Lee}},\ and\ \bibinfo {author} {\bibfnamefont {Y.-S.}\
  \bibnamefont {Choi}},\ }\href {https://doi.org/10.1039/D2CP04542G} {\bibfield
   {journal} {\bibinfo  {journal} {Physical Chemistry Chemical Physics}\
  }\textbf {\bibinfo {volume} {24}},\ \bibinfo {pages} {26870} (\bibinfo {year}
  {2022})}\BibitemShut {NoStop}%
\bibitem [{\citenamefont {Bånkestad}\ \emph {et~al.}(2024)\citenamefont
  {Bånkestad}, \citenamefont {Dorst}, \citenamefont {Widmalm},\ and\
  \citenamefont {Rönnols}}]{bankestad_carbohydrate_2024}%
  \BibitemOpen
  \bibfield  {author} {\bibinfo {author} {\bibfnamefont {M.}~\bibnamefont
  {Bånkestad}}, \bibinfo {author} {\bibfnamefont {K.~M.}\ \bibnamefont
  {Dorst}}, \bibinfo {author} {\bibfnamefont {G.}~\bibnamefont {Widmalm}},\
  and\ \bibinfo {author} {\bibfnamefont {J.}~\bibnamefont {Rönnols}},\ }\href
  {https://doi.org/10.1039/D4RA03428G} {\bibfield  {journal} {\bibinfo
  {journal} {RSC Advances}\ }\textbf {\bibinfo {volume} {14}},\ \bibinfo
  {pages} {26585} (\bibinfo {year} {2024})}\BibitemShut {NoStop}%
\bibitem [{\citenamefont {Rupp}, \citenamefont {Ramakrishnan},\ and\
  \citenamefont {Von~Lilienfeld}(2015)}]{rupp_machine_2015}%
  \BibitemOpen
  \bibfield  {author} {\bibinfo {author} {\bibfnamefont {M.}~\bibnamefont
  {Rupp}}, \bibinfo {author} {\bibfnamefont {R.}~\bibnamefont {Ramakrishnan}},\
  and\ \bibinfo {author} {\bibfnamefont {O.~A.}\ \bibnamefont
  {Von~Lilienfeld}},\ }\href {https://doi.org/10.1021/acs.jpclett.5b01456}
  {\bibfield  {journal} {\bibinfo  {journal} {The Journal of Physical Chemistry
  Letters}\ }\textbf {\bibinfo {volume} {6}},\ \bibinfo {pages} {3309}
  (\bibinfo {year} {2015})}\BibitemShut {NoStop}%
\bibitem [{\citenamefont {Paruzzo}\ \emph {et~al.}(2018)\citenamefont
  {Paruzzo}, \citenamefont {Hofstetter}, \citenamefont {Musil}, \citenamefont
  {De}, \citenamefont {Ceriotti},\ and\ \citenamefont
  {Emsley}}]{paruzzo_chemical_2018}%
  \BibitemOpen
  \bibfield  {author} {\bibinfo {author} {\bibfnamefont {F.~M.}\ \bibnamefont
  {Paruzzo}}, \bibinfo {author} {\bibfnamefont {A.}~\bibnamefont {Hofstetter}},
  \bibinfo {author} {\bibfnamefont {F.}~\bibnamefont {Musil}}, \bibinfo
  {author} {\bibfnamefont {S.}~\bibnamefont {De}}, \bibinfo {author}
  {\bibfnamefont {M.}~\bibnamefont {Ceriotti}},\ and\ \bibinfo {author}
  {\bibfnamefont {L.}~\bibnamefont {Emsley}},\ }\href
  {https://doi.org/10.1038/s41467-018-06972-x} {\bibfield  {journal} {\bibinfo
  {journal} {Nature Communications}\ }\textbf {\bibinfo {volume} {9}},\
  \bibinfo {pages} {4501} (\bibinfo {year} {2018})}\BibitemShut {NoStop}%
\bibitem [{\citenamefont {Chaker}\ \emph {et~al.}(2019)\citenamefont {Chaker},
  \citenamefont {Salanne}, \citenamefont {Delaye},\ and\ \citenamefont
  {Charpentier}}]{chaker_nmr_2019}%
  \BibitemOpen
  \bibfield  {author} {\bibinfo {author} {\bibfnamefont {Z.}~\bibnamefont
  {Chaker}}, \bibinfo {author} {\bibfnamefont {M.}~\bibnamefont {Salanne}},
  \bibinfo {author} {\bibfnamefont {J.-M.}\ \bibnamefont {Delaye}},\ and\
  \bibinfo {author} {\bibfnamefont {T.}~\bibnamefont {Charpentier}},\ }\href
  {https://doi.org/10.1039/C9CP02803J} {\bibfield  {journal} {\bibinfo
  {journal} {Physical Chemistry Chemical Physics}\ }\textbf {\bibinfo {volume}
  {21}},\ \bibinfo {pages} {21709} (\bibinfo {year} {2019})}\BibitemShut
  {NoStop}%
\bibitem [{\citenamefont {Venetos}, \citenamefont {Wen},\ and\ \citenamefont
  {Persson}(2023)}]{venetos_machine_2023}%
  \BibitemOpen
  \bibfield  {author} {\bibinfo {author} {\bibfnamefont {M.~C.}\ \bibnamefont
  {Venetos}}, \bibinfo {author} {\bibfnamefont {M.}~\bibnamefont {Wen}},\ and\
  \bibinfo {author} {\bibfnamefont {K.~A.}\ \bibnamefont {Persson}},\ }\href
  {https://doi.org/10.1021/acs.jpca.2c07530} {\bibfield  {journal} {\bibinfo
  {journal} {The Journal of Physical Chemistry A}\ }\textbf {\bibinfo {volume}
  {127}},\ \bibinfo {pages} {2388} (\bibinfo {year} {2023})}\BibitemShut
  {NoStop}%
\bibitem [{\citenamefont
  {Charpentier}(2024)}]{charpentier_first-principles_2024}%
  \BibitemOpen
  \bibfield  {author} {\bibinfo {author} {\bibfnamefont {T.}~\bibnamefont
  {Charpentier}},\ }\href {https://doi.org/10.1039/D4FD00129J} {\bibfield
  {journal} {\bibinfo  {journal} {Faraday Discussions}\ } (\bibinfo {year}
  {2024}),\ 10.1039/D4FD00129J},\ \bibinfo {note} {publisher: The Royal Society
  of Chemistry}\BibitemShut {NoStop}%
\bibitem [{\citenamefont {Harper}\ \emph {et~al.}(2024)\citenamefont {Harper},
  \citenamefont {Köcher}, \citenamefont {Reuter},\ and\ \citenamefont
  {Scheurer}}]{harper_performance_2024}%
  \BibitemOpen
  \bibfield  {author} {\bibinfo {author} {\bibfnamefont {A.~F.}\ \bibnamefont
  {Harper}}, \bibinfo {author} {\bibfnamefont {S.~S.}\ \bibnamefont {Köcher}},
  \bibinfo {author} {\bibfnamefont {K.}~\bibnamefont {Reuter}},\ and\ \bibinfo
  {author} {\bibfnamefont {C.}~\bibnamefont {Scheurer}},\ }\href
  {https://doi.org/10.26434/chemrxiv-2024-j0kp2} {\enquote {\bibinfo {title}
  {Performance metrics for tensorial learning: prediction of {Li4Ti5O12}
  nuclear magnetic resonance observables at experimental accuracy},}\ }
  (\bibinfo {year} {2024}),\ \bibinfo {note}
  {10.26434/chemrxiv-2024-j0kp2}\BibitemShut {NoStop}%
\bibitem [{\citenamefont {Batzner}\ \emph {et~al.}(2022)\citenamefont
  {Batzner}, \citenamefont {Musaelian}, \citenamefont {Sun}, \citenamefont
  {Geiger}, \citenamefont {Mailoa}, \citenamefont {Kornbluth}, \citenamefont
  {Molinari}, \citenamefont {Smidt},\ and\ \citenamefont
  {Kozinsky}}]{batzner_e3-equivariant_2022}%
  \BibitemOpen
  \bibfield  {author} {\bibinfo {author} {\bibfnamefont {S.}~\bibnamefont
  {Batzner}}, \bibinfo {author} {\bibfnamefont {A.}~\bibnamefont {Musaelian}},
  \bibinfo {author} {\bibfnamefont {L.}~\bibnamefont {Sun}}, \bibinfo {author}
  {\bibfnamefont {M.}~\bibnamefont {Geiger}}, \bibinfo {author} {\bibfnamefont
  {J.~P.}\ \bibnamefont {Mailoa}}, \bibinfo {author} {\bibfnamefont
  {M.}~\bibnamefont {Kornbluth}}, \bibinfo {author} {\bibfnamefont
  {N.}~\bibnamefont {Molinari}}, \bibinfo {author} {\bibfnamefont {T.~E.}\
  \bibnamefont {Smidt}},\ and\ \bibinfo {author} {\bibfnamefont
  {B.}~\bibnamefont {Kozinsky}},\ }\href
  {https://doi.org/10.1038/s41467-022-29939-5} {\bibfield  {journal} {\bibinfo
  {journal} {Nature Communications}\ }\textbf {\bibinfo {volume} {13}},\
  \bibinfo {pages} {2453} (\bibinfo {year} {2022})}\BibitemShut {NoStop}%
\bibitem [{\citenamefont {Haeberlen}(1976)}]{haeberlen_high_1976}%
  \BibitemOpen
  \bibfield  {author} {\bibinfo {author} {\bibfnamefont {U.}~\bibnamefont
  {Haeberlen}},\ }\href@noop {} {\emph {\bibinfo {title} {High resolution {NMR}
  in solids: selective averaging}}},\ \bibinfo {series} {Advances in magnetic
  resonance : {Supplement}}\ No.~\bibinfo {number} {1}\ (\bibinfo  {publisher}
  {Academic Press},\ \bibinfo {address} {New York},\ \bibinfo {year}
  {1976})\BibitemShut {NoStop}%
\bibitem [{\citenamefont {Harris}\ \emph {et~al.}(2008)\citenamefont {Harris},
  \citenamefont {Becker}, \citenamefont {Cabral De~Menezes}, \citenamefont
  {Granger}, \citenamefont {Hoffman},\ and\ \citenamefont
  {Zilm}}]{harris_further_2008}%
  \BibitemOpen
  \bibfield  {author} {\bibinfo {author} {\bibfnamefont {R.~K.}\ \bibnamefont
  {Harris}}, \bibinfo {author} {\bibfnamefont {E.~D.}\ \bibnamefont {Becker}},
  \bibinfo {author} {\bibfnamefont {S.~M.}\ \bibnamefont {Cabral De~Menezes}},
  \bibinfo {author} {\bibfnamefont {P.}~\bibnamefont {Granger}}, \bibinfo
  {author} {\bibfnamefont {R.~E.}\ \bibnamefont {Hoffman}},\ and\ \bibinfo
  {author} {\bibfnamefont {K.~W.}\ \bibnamefont {Zilm}},\ }\href
  {https://doi.org/10.1016/j.ssnmr.2008.02.004} {\bibfield  {journal} {\bibinfo
   {journal} {Solid State Nuclear Magnetic Resonance}\ }\textbf {\bibinfo
  {volume} {33}},\ \bibinfo {pages} {41} (\bibinfo {year} {2008})}\BibitemShut
  {NoStop}%
\bibitem [{\citenamefont {Mason}(1993)}]{mason_conventions_1993}%
  \BibitemOpen
  \bibfield  {author} {\bibinfo {author} {\bibfnamefont {J.}~\bibnamefont
  {Mason}},\ }\href {https://doi.org/10.1016/0926-2040(93)90010-K} {\bibfield
  {journal} {\bibinfo  {journal} {Solid State Nuclear Magnetic Resonance}\
  }\textbf {\bibinfo {volume} {2}},\ \bibinfo {pages} {285} (\bibinfo {year}
  {1993})}\BibitemShut {NoStop}%
\bibitem [{noa(1976)}]{noauthor_advances_1976}%
  \BibitemOpen
  in\ \href {https://doi.org/10.1016/B978-0-12-025561-0.50001-0}
  {{\emph {\bibinfo {booktitle} {High {Resolution} {Nmr} in
  {Solids} {Selective} {Averaging}}}}}\ (\bibinfo  {publisher} {Elsevier},\
  \bibinfo {year} {1976})\ p.~\bibinfo {pages} {ii}\BibitemShut {NoStop}%
\bibitem [{\citenamefont {Thomas}\ \emph {et~al.}(2018)\citenamefont {Thomas},
  \citenamefont {Smidt}, \citenamefont {Kearnes}, \citenamefont {Yang},
  \citenamefont {Li}, \citenamefont {Kohlhoff},\ and\ \citenamefont
  {Riley}}]{thomas_tensor_2018}%
  \BibitemOpen
  \bibfield  {author} {\bibinfo {author} {\bibfnamefont {N.}~\bibnamefont
  {Thomas}}, \bibinfo {author} {\bibfnamefont {T.}~\bibnamefont {Smidt}},
  \bibinfo {author} {\bibfnamefont {S.}~\bibnamefont {Kearnes}}, \bibinfo
  {author} {\bibfnamefont {L.}~\bibnamefont {Yang}}, \bibinfo {author}
  {\bibfnamefont {L.}~\bibnamefont {Li}}, \bibinfo {author} {\bibfnamefont
  {K.}~\bibnamefont {Kohlhoff}},\ and\ \bibinfo {author} {\bibfnamefont
  {P.}~\bibnamefont {Riley}},\ }\href
  {https://doi.org/10.48550/ARXIV.1802.08219} {\enquote {\bibinfo {title}
  {Tensor field networks: {Rotation}- and translation-equivariant neural
  networks for {3D} point clouds},}\ } (\bibinfo {year} {2018}),\ \bibinfo
  {note} {version Number: 3}\BibitemShut {NoStop}%
\bibitem [{\citenamefont {Bonhomme}\ \emph {et~al.}(2012)\citenamefont
  {Bonhomme}, \citenamefont {Gervais}, \citenamefont {Babonneau}, \citenamefont
  {Coelho}, \citenamefont {Pourpoint}, \citenamefont {Azaïs}, \citenamefont
  {Ashbrook}, \citenamefont {Griffin}, \citenamefont {Yates}, \citenamefont
  {Mauri},\ and\ \citenamefont {Pickard}}]{bonhomme_first-principles_2012}%
  \BibitemOpen
  \bibfield  {author} {\bibinfo {author} {\bibfnamefont {C.}~\bibnamefont
  {Bonhomme}}, \bibinfo {author} {\bibfnamefont {C.}~\bibnamefont {Gervais}},
  \bibinfo {author} {\bibfnamefont {F.}~\bibnamefont {Babonneau}}, \bibinfo
  {author} {\bibfnamefont {C.}~\bibnamefont {Coelho}}, \bibinfo {author}
  {\bibfnamefont {F.}~\bibnamefont {Pourpoint}}, \bibinfo {author}
  {\bibfnamefont {T.}~\bibnamefont {Azaïs}}, \bibinfo {author} {\bibfnamefont
  {S.~E.}\ \bibnamefont {Ashbrook}}, \bibinfo {author} {\bibfnamefont {J.~M.}\
  \bibnamefont {Griffin}}, \bibinfo {author} {\bibfnamefont {J.~R.}\
  \bibnamefont {Yates}}, \bibinfo {author} {\bibfnamefont {F.}~\bibnamefont
  {Mauri}},\ and\ \bibinfo {author} {\bibfnamefont {C.~J.}\ \bibnamefont
  {Pickard}},\ }\href {https://doi.org/10.1021/cr300108a} {\bibfield  {journal}
  {\bibinfo  {journal} {Chemical Reviews}\ }\textbf {\bibinfo {volume} {112}},\
  \bibinfo {pages} {5733} (\bibinfo {year} {2012})}\BibitemShut {NoStop}%
\bibitem [{\citenamefont {Anet}\ \emph {et~al.}(1990)\citenamefont {Anet},
  \citenamefont {O'Leary}, \citenamefont {Wade},\ and\ \citenamefont
  {Johnson}}]{anet_nmr_1990}%
  \BibitemOpen
  \bibfield  {author} {\bibinfo {author} {\bibfnamefont {F.~A.}\ \bibnamefont
  {Anet}}, \bibinfo {author} {\bibfnamefont {D.~J.}\ \bibnamefont {O'Leary}},
  \bibinfo {author} {\bibfnamefont {C.~G.}\ \bibnamefont {Wade}},\ and\
  \bibinfo {author} {\bibfnamefont {R.~D.}\ \bibnamefont {Johnson}},\ }\href
  {https://doi.org/10.1016/0009-2614(90)85237-7} {\bibfield  {journal}
  {\bibinfo  {journal} {Chemical Physics Letters}\ }\textbf {\bibinfo {volume}
  {171}},\ \bibinfo {pages} {401} (\bibinfo {year} {1990})}\BibitemShut
  {NoStop}%
\bibitem [{\citenamefont {Geiger}\ and\ \citenamefont
  {Smidt}(2022)}]{geiger_e3nn_2022}%
  \BibitemOpen
  \bibfield  {author} {\bibinfo {author} {\bibfnamefont {M.}~\bibnamefont
  {Geiger}}\ and\ \bibinfo {author} {\bibfnamefont {T.}~\bibnamefont {Smidt}},\
  }\href {https://doi.org/10.48550/ARXIV.2207.09453} {\enquote {\bibinfo
  {title} {e3nn: {Euclidean} {Neural} {Networks}},}\ } (\bibinfo {year}
  {2022}),\ \bibinfo {note} {version Number: 1}\BibitemShut {NoStop}%
\bibitem [{\citenamefont {Thompson}\ \emph {et~al.}(2022)\citenamefont
  {Thompson}, \citenamefont {Aktulga}, \citenamefont {Berger}, \citenamefont
  {Bolintineanu}, \citenamefont {Brown}, \citenamefont {Crozier}, \citenamefont
  {in~'t Veld}, \citenamefont {Kohlmeyer}, \citenamefont {Moore}, \citenamefont
  {Nguyen}, \citenamefont {Shan}, \citenamefont {Stevens}, \citenamefont
  {Tranchida}, \citenamefont {Trott},\ and\ \citenamefont
  {Plimpton}}]{thompson_lammps_2022}%
  \BibitemOpen
  \bibfield  {author} {\bibinfo {author} {\bibfnamefont {A.~P.}\ \bibnamefont
  {Thompson}}, \bibinfo {author} {\bibfnamefont {H.~M.}\ \bibnamefont
  {Aktulga}}, \bibinfo {author} {\bibfnamefont {R.}~\bibnamefont {Berger}},
  \bibinfo {author} {\bibfnamefont {D.~S.}\ \bibnamefont {Bolintineanu}},
  \bibinfo {author} {\bibfnamefont {W.~M.}\ \bibnamefont {Brown}}, \bibinfo
  {author} {\bibfnamefont {P.~S.}\ \bibnamefont {Crozier}}, \bibinfo {author}
  {\bibfnamefont {P.~J.}\ \bibnamefont {in~'t Veld}}, \bibinfo {author}
  {\bibfnamefont {A.}~\bibnamefont {Kohlmeyer}}, \bibinfo {author}
  {\bibfnamefont {S.~G.}\ \bibnamefont {Moore}}, \bibinfo {author}
  {\bibfnamefont {T.~D.}\ \bibnamefont {Nguyen}}, \bibinfo {author}
  {\bibfnamefont {R.}~\bibnamefont {Shan}}, \bibinfo {author} {\bibfnamefont
  {M.~J.}\ \bibnamefont {Stevens}}, \bibinfo {author} {\bibfnamefont
  {J.}~\bibnamefont {Tranchida}}, \bibinfo {author} {\bibfnamefont
  {C.}~\bibnamefont {Trott}},\ and\ \bibinfo {author} {\bibfnamefont {S.~J.}\
  \bibnamefont {Plimpton}},\ }\href {https://doi.org/10.1016/j.cpc.2021.108171}
  {\bibfield  {journal} {\bibinfo  {journal} {Computer Physics Communications}\
  }\textbf {\bibinfo {volume} {271}},\ \bibinfo {pages} {108171} (\bibinfo
  {year} {2022})}\BibitemShut {NoStop}%
\bibitem [{\citenamefont {Erhard}\ \emph {et~al.}(2022)\citenamefont {Erhard},
  \citenamefont {Rohrer}, \citenamefont {Albe},\ and\ \citenamefont
  {Deringer}}]{erhard_machine-learned_2022}%
  \BibitemOpen
  \bibfield  {author} {\bibinfo {author} {\bibfnamefont {L.~C.}\ \bibnamefont
  {Erhard}}, \bibinfo {author} {\bibfnamefont {J.}~\bibnamefont {Rohrer}},
  \bibinfo {author} {\bibfnamefont {K.}~\bibnamefont {Albe}},\ and\ \bibinfo
  {author} {\bibfnamefont {V.~L.}\ \bibnamefont {Deringer}},\ }\href
  {https://doi.org/10.1038/s41524-022-00768-w} {\bibfield  {journal} {\bibinfo
  {journal} {npj Computational Materials}\ }\textbf {\bibinfo {volume} {8}},\
  \bibinfo {pages} {90} (\bibinfo {year} {2022})}\BibitemShut {NoStop}%
\bibitem [{\citenamefont {Carré}\ \emph {et~al.}(2008)\citenamefont {Carré},
  \citenamefont {Horbach}, \citenamefont {Ispas},\ and\ \citenamefont
  {Kob}}]{carre_new_2008}%
  \BibitemOpen
  \bibfield  {author} {\bibinfo {author} {\bibfnamefont {A.}~\bibnamefont
  {Carré}}, \bibinfo {author} {\bibfnamefont {J.}~\bibnamefont {Horbach}},
  \bibinfo {author} {\bibfnamefont {S.}~\bibnamefont {Ispas}},\ and\ \bibinfo
  {author} {\bibfnamefont {W.}~\bibnamefont {Kob}},\ }\href
  {https://doi.org/10.1209/0295-5075/82/17001} {\bibfield  {journal} {\bibinfo
  {journal} {EPL (Europhysics Letters)}\ }\textbf {\bibinfo {volume} {82}},\
  \bibinfo {pages} {17001} (\bibinfo {year} {2008})}\BibitemShut {NoStop}%
\bibitem [{\citenamefont {Profeta}, \citenamefont {Mauri},\ and\ \citenamefont
  {Pickard}(2003)}]{profeta_accurate_2003}%
  \BibitemOpen
  \bibfield  {author} {\bibinfo {author} {\bibfnamefont {M.}~\bibnamefont
  {Profeta}}, \bibinfo {author} {\bibfnamefont {F.}~\bibnamefont {Mauri}},\
  and\ \bibinfo {author} {\bibfnamefont {C.~J.}\ \bibnamefont {Pickard}},\
  }\href {https://doi.org/10.1021/ja027124r} {\bibfield  {journal} {\bibinfo
  {journal} {Journal of the American Chemical Society}\ }\textbf {\bibinfo
  {volume} {125}},\ \bibinfo {pages} {541} (\bibinfo {year}
  {2003})}\BibitemShut {NoStop}%
\bibitem [{\citenamefont {Clark}\ \emph {et~al.}(2005)\citenamefont {Clark},
  \citenamefont {Segall}, \citenamefont {Pickard}, \citenamefont {Hasnip},
  \citenamefont {Probert}, \citenamefont {Refson},\ and\ \citenamefont
  {Payne}}]{clark_first_2005}%
  \BibitemOpen
  \bibfield  {author} {\bibinfo {author} {\bibfnamefont {S.~J.}\ \bibnamefont
  {Clark}}, \bibinfo {author} {\bibfnamefont {M.~D.}\ \bibnamefont {Segall}},
  \bibinfo {author} {\bibfnamefont {C.~J.}\ \bibnamefont {Pickard}}, \bibinfo
  {author} {\bibfnamefont {P.~J.}\ \bibnamefont {Hasnip}}, \bibinfo {author}
  {\bibfnamefont {M.~I.~J.}\ \bibnamefont {Probert}}, \bibinfo {author}
  {\bibfnamefont {K.}~\bibnamefont {Refson}},\ and\ \bibinfo {author}
  {\bibfnamefont {M.~C.}\ \bibnamefont {Payne}},\ }\href
  {https://doi.org/10.1524/zkri.220.5.567.65075} {\bibfield  {journal}
  {\bibinfo  {journal} {Zeitschrift für Kristallographie - Crystalline
  Materials}\ }\textbf {\bibinfo {volume} {220}},\ \bibinfo {pages} {567}
  (\bibinfo {year} {2005})}\BibitemShut {NoStop}%
\bibitem [{\citenamefont {Perdew}, \citenamefont {Burke},\ and\ \citenamefont
  {Ernzerhof}(1996)}]{perdew_generalized_1996}%
  \BibitemOpen
  \bibfield  {author} {\bibinfo {author} {\bibfnamefont {J.~P.}\ \bibnamefont
  {Perdew}}, \bibinfo {author} {\bibfnamefont {K.}~\bibnamefont {Burke}},\ and\
  \bibinfo {author} {\bibfnamefont {M.}~\bibnamefont {Ernzerhof}},\ }\href
  {https://doi.org/10/bppfwt} {\bibfield  {journal} {\bibinfo  {journal}
  {Physical Review Letters}\ }\textbf {\bibinfo {volume} {77}},\ \bibinfo
  {pages} {3865} (\bibinfo {year} {1996})},\ \bibinfo {note} {publisher:
  American Physical Society}\BibitemShut {NoStop}%
\bibitem [{\citenamefont {Thomas Du~Toit}\ and\ \citenamefont
  {Deringer}(2023)}]{thomas_du_toit_cross-platform_2023}%
  \BibitemOpen
  \bibfield  {author} {\bibinfo {author} {\bibfnamefont {D.~F.}\ \bibnamefont
  {Thomas Du~Toit}}\ and\ \bibinfo {author} {\bibfnamefont {V.~L.}\
  \bibnamefont {Deringer}},\ }\href {https://doi.org/10.1063/5.0155618}
  {\bibfield  {journal} {\bibinfo  {journal} {The Journal of Chemical Physics}\
  }\textbf {\bibinfo {volume} {159}},\ \bibinfo {pages} {024803} (\bibinfo
  {year} {2023})}\BibitemShut {NoStop}%
\bibitem [{\citenamefont {Svenningsson}\ and\ \citenamefont
  {Mueller}(2023)}]{svenningsson_tensorview_2023}%
  \BibitemOpen
  \bibfield  {author} {\bibinfo {author} {\bibfnamefont {L.}~\bibnamefont
  {Svenningsson}}\ and\ \bibinfo {author} {\bibfnamefont {L.~J.}\ \bibnamefont
  {Mueller}},\ }\href {https://doi.org/10.1016/j.ssnmr.2022.101849} {\bibfield
  {journal} {\bibinfo  {journal} {Solid State Nuclear Magnetic Resonance}\
  }\textbf {\bibinfo {volume} {123}},\ \bibinfo {pages} {101849} (\bibinfo
  {year} {2023})}\BibitemShut {NoStop}%
\bibitem [{\citenamefont {Shankar}(2013)}]{shankar_principles_2013}%
  \BibitemOpen
  \bibfield  {author} {\bibinfo {author} {\bibfnamefont {R.}~\bibnamefont
  {Shankar}},\ }\href@noop {} {\emph {\bibinfo {title} {Principles of quantum
  mechanics}}}\ (\bibinfo  {publisher} {Springer},\ \bibinfo {year}
  {2013})\BibitemShut {NoStop}%
\bibitem [{\citenamefont {Erhard}\ \emph
  {et~al.}(2024{\natexlab{a}})\citenamefont {Erhard}, \citenamefont {Rohrer},
  \citenamefont {Albe},\ and\ \citenamefont
  {Deringer}}]{erhard_modelling_2024}%
  \BibitemOpen
  \bibfield  {author} {\bibinfo {author} {\bibfnamefont {L.~C.}\ \bibnamefont
  {Erhard}}, \bibinfo {author} {\bibfnamefont {J.}~\bibnamefont {Rohrer}},
  \bibinfo {author} {\bibfnamefont {K.}~\bibnamefont {Albe}},\ and\ \bibinfo
  {author} {\bibfnamefont {V.~L.}\ \bibnamefont {Deringer}},\ }\href
  {https://doi.org/10.1038/s41467-024-45840-9} {\bibfield  {journal} {\bibinfo
  {journal} {Nature Communications}\ }\textbf {\bibinfo {volume} {15}},\
  \bibinfo {pages} {1927} (\bibinfo {year} {2024}{\natexlab{a}})}\BibitemShut
  {NoStop}%
\bibitem [{noa()}]{noauthor_soprano_nodate}%
  \BibitemOpen
  \href {https://github.com/CCP-NC/soprano} {\enquote {\bibinfo {title}
  {Soprano},}\ }\BibitemShut {NoStop}%
\bibitem [{\citenamefont {Stukowski}(2010)}]{stukowski_visualization_2010}%
  \BibitemOpen
  \bibfield  {author} {\bibinfo {author} {\bibfnamefont {A.}~\bibnamefont
  {Stukowski}},\ }\href {https://doi.org/10.1088/0965-0393/18/1/015012}
  {\bibfield  {journal} {\bibinfo  {journal} {Modelling and Simulation in
  Materials Science and Engineering}\ }\textbf {\bibinfo {volume} {18}},\
  \bibinfo {pages} {015012} (\bibinfo {year} {2010})}\BibitemShut {NoStop}%
\bibitem [{\citenamefont {Baerlocher}\ \emph {et~al.}()\citenamefont
  {Baerlocher}, \citenamefont {Brouwer}, \citenamefont {Marler},\ and\
  \citenamefont {McCusker}}]{baerlocher_database_nodate}%
  \BibitemOpen
  \bibfield  {author} {\bibinfo {author} {\bibfnamefont {C.}~\bibnamefont
  {Baerlocher}}, \bibinfo {author} {\bibfnamefont {D.}~\bibnamefont {Brouwer}},
  \bibinfo {author} {\bibfnamefont {B.}~\bibnamefont {Marler}},\ and\ \bibinfo
  {author} {\bibfnamefont {L.}~\bibnamefont {McCusker}},\ }\href
  {https://www.iza-structure.org/databases/} {\enquote {\bibinfo {title}
  {Database of {Zeolite} {Structures}},}\ }\BibitemShut {NoStop}%
\bibitem [{\citenamefont {Erhard}\ \emph
  {et~al.}(2024{\natexlab{b}})\citenamefont {Erhard}, \citenamefont {Utt},
  \citenamefont {Klomp},\ and\ \citenamefont {Albe}}]{erhard_crystal_2024}%
  \BibitemOpen
  \bibfield  {author} {\bibinfo {author} {\bibfnamefont {L.~C.}\ \bibnamefont
  {Erhard}}, \bibinfo {author} {\bibfnamefont {D.}~\bibnamefont {Utt}},
  \bibinfo {author} {\bibfnamefont {A.~J.}\ \bibnamefont {Klomp}},\ and\
  \bibinfo {author} {\bibfnamefont {K.}~\bibnamefont {Albe}},\ }\href
  {https://doi.org/10.1088/1361-651X/ad64f3} {\bibfield  {journal} {\bibinfo
  {journal} {Modelling and Simulation in Materials Science and Engineering}\
  }\textbf {\bibinfo {volume} {32}},\ \bibinfo {pages} {065029} (\bibinfo
  {year} {2024}{\natexlab{b}})}\BibitemShut {NoStop}%
\bibitem [{\citenamefont {Walker}\ \emph {et~al.}(1958)\citenamefont {Walker},
  \citenamefont {Zerfoss}, \citenamefont {Holley},\ and\ \citenamefont
  {Gross}}]{walker_temperature_1958}%
  \BibitemOpen
  \bibfield  {author} {\bibinfo {author} {\bibfnamefont {R.}~\bibnamefont
  {Walker}}, \bibinfo {author} {\bibfnamefont {S.}~\bibnamefont {Zerfoss}},
  \bibinfo {author} {\bibfnamefont {S.}~\bibnamefont {Holley}},\ and\ \bibinfo
  {author} {\bibfnamefont {L.}~\bibnamefont {Gross}},\ }\href
  {https://doi.org/10.6028/jres.061.026} {\bibfield  {journal} {\bibinfo
  {journal} {Journal of Research of the National Bureau of Standards}\ }\textbf
  {\bibinfo {volume} {61}},\ \bibinfo {pages} {251} (\bibinfo {year}
  {1958})}\BibitemShut {NoStop}%
\bibitem [{pea(1973)}]{peacor_high-temperature_1973}%
  \BibitemOpen
  \href {https://doi.org/10.1524/zkri.1973.138.1-4.274} {\bibfield  {journal}
  {\bibinfo  {journal} {Zeitschrift für Kristallographie - Crystalline
  Materials}\ }\textbf {\bibinfo {volume} {138}},\ \bibinfo {pages} {274}
  (\bibinfo {year} {1973})}\BibitemShut {NoStop}%
\bibitem [{\citenamefont {Wright}\ and\ \citenamefont
  {Leadbetter}(1975)}]{wright_structures_1975}%
  \BibitemOpen
  \bibfield  {author} {\bibinfo {author} {\bibfnamefont {A.~F.}\ \bibnamefont
  {Wright}}\ and\ \bibinfo {author} {\bibfnamefont {A.~J.}\ \bibnamefont
  {Leadbetter}},\ }\href {https://doi.org/10.1080/00318087508228690} {\bibfield
   {journal} {\bibinfo  {journal} {Philosophical Magazine}\ }\textbf {\bibinfo
  {volume} {31}},\ \bibinfo {pages} {1391} (\bibinfo {year}
  {1975})}\BibitemShut {NoStop}%
\bibitem [{\citenamefont {Hatch}\ and\ \citenamefont
  {Ghose}(1991)}]{hatch_alpha--beta_1991}%
  \BibitemOpen
  \bibfield  {author} {\bibinfo {author} {\bibfnamefont {D.}~\bibnamefont
  {Hatch}}\ and\ \bibinfo {author} {\bibfnamefont {S.}~\bibnamefont {Ghose}},\
  }\href {https://doi.org/10.1007/BF00202234} {\bibfield  {journal} {\bibinfo
  {journal} {Physics and Chemistry of Minerals}\ }\textbf {\bibinfo {volume}
  {17}},\ \bibinfo {pages} {554} (\bibinfo {year} {1991})}\BibitemShut
  {NoStop}%
\bibitem [{\citenamefont {Yuan}\ and\ \citenamefont
  {Huang}(2012)}]{yuan__2012}%
  \BibitemOpen
  \bibfield  {author} {\bibinfo {author} {\bibfnamefont {F.}~\bibnamefont
  {Yuan}}\ and\ \bibinfo {author} {\bibfnamefont {L.}~\bibnamefont {Huang}},\
  }\href {https://doi.org/10.1103/PhysRevB.85.134114} {\bibfield  {journal}
  {\bibinfo  {journal} {Physical Review B}\ }\textbf {\bibinfo {volume} {85}},\
  \bibinfo {pages} {134114} (\bibinfo {year} {2012})}\BibitemShut {NoStop}%
\bibitem [{\citenamefont {Sun}, \citenamefont {Ruzsinszky},\ and\ \citenamefont
  {Perdew}(2015)}]{sun_strongly_2015}%
  \BibitemOpen
  \bibfield  {author} {\bibinfo {author} {\bibfnamefont {J.}~\bibnamefont
  {Sun}}, \bibinfo {author} {\bibfnamefont {A.}~\bibnamefont {Ruzsinszky}},\
  and\ \bibinfo {author} {\bibfnamefont {J.}~\bibnamefont {Perdew}},\ }\href
  {https://doi.org/10.1103/PhysRevLett.115.036402} {\bibfield  {journal}
  {\bibinfo  {journal} {Physical Review Letters}\ }\textbf {\bibinfo {volume}
  {115}},\ \bibinfo {pages} {036402} (\bibinfo {year} {2015})}\BibitemShut
  {NoStop}%
\bibitem [{\citenamefont {Ashbrook}, \citenamefont {Dawson},\ and\
  \citenamefont {Seymour}(2014)}]{ashbrook_recent_2014}%
  \BibitemOpen
  \bibfield  {author} {\bibinfo {author} {\bibfnamefont {S.~E.}\ \bibnamefont
  {Ashbrook}}, \bibinfo {author} {\bibfnamefont {D.~M.}\ \bibnamefont
  {Dawson}},\ and\ \bibinfo {author} {\bibfnamefont {V.~R.}\ \bibnamefont
  {Seymour}},\ }\href {https://doi.org/10.1039/C4CP00578C} {\bibfield
  {journal} {\bibinfo  {journal} {Phys. Chem. Chem. Phys.}\ }\textbf {\bibinfo
  {volume} {16}},\ \bibinfo {pages} {8223} (\bibinfo {year}
  {2014})}\BibitemShut {NoStop}%
\bibitem [{\citenamefont {G.M.~Rankin}\ \emph {et~al.}(2023)\citenamefont
  {G.M.~Rankin}, \citenamefont {Pourpoint}, \citenamefont {Duong},
  \citenamefont {Delevoye}, \citenamefont {Amoureux},\ and\ \citenamefont
  {Lafon}}]{gm_rankin_advances_2023}%
  \BibitemOpen
  \bibfield  {author} {\bibinfo {author} {\bibfnamefont {A.}~\bibnamefont
  {G.M.~Rankin}}, \bibinfo {author} {\bibfnamefont {F.}~\bibnamefont
  {Pourpoint}}, \bibinfo {author} {\bibfnamefont {N.~T.}\ \bibnamefont
  {Duong}}, \bibinfo {author} {\bibfnamefont {L.}~\bibnamefont {Delevoye}},
  \bibinfo {author} {\bibfnamefont {J.-P.}\ \bibnamefont {Amoureux}},\ and\
  \bibinfo {author} {\bibfnamefont {O.}~\bibnamefont {Lafon}},\ }in\ \href
  {https://doi.org/10.1016/B978-0-12-823144-9.00192-8} {{\emph {\bibinfo {booktitle} {Comprehensive Inorganic Chemistry
  III}}}}\ (\bibinfo  {publisher} {Elsevier},\ \bibinfo {year} {2023})\ pp.\
  \bibinfo {pages} {534--582}\BibitemShut {NoStop}%
\bibitem [{\citenamefont {Trillot}\ \emph {et~al.}(2024)\citenamefont
  {Trillot}, \citenamefont {Lam}, \citenamefont {Ispas}, \citenamefont {Kandy},
  \citenamefont {Tuckerman}, \citenamefont {Tarrat},\ and\ \citenamefont
  {Benoit}}]{trillot_elaboration_2024}%
  \BibitemOpen
  \bibfield  {author} {\bibinfo {author} {\bibfnamefont {S.}~\bibnamefont
  {Trillot}}, \bibinfo {author} {\bibfnamefont {J.}~\bibnamefont {Lam}},
  \bibinfo {author} {\bibfnamefont {S.}~\bibnamefont {Ispas}}, \bibinfo
  {author} {\bibfnamefont {A.~K.~A.}\ \bibnamefont {Kandy}}, \bibinfo {author}
  {\bibfnamefont {M.~E.}\ \bibnamefont {Tuckerman}}, \bibinfo {author}
  {\bibfnamefont {N.}~\bibnamefont {Tarrat}},\ and\ \bibinfo {author}
  {\bibfnamefont {M.}~\bibnamefont {Benoit}},\ }\href
  {https://doi.org/10.1016/j.commatsci.2024.112848} {\bibfield  {journal}
  {\bibinfo  {journal} {Computational Materials Science}\ }\textbf {\bibinfo
  {volume} {236}},\ \bibinfo {pages} {112848} (\bibinfo {year}
  {2024})}\BibitemShut {NoStop}%
\bibitem [{\citenamefont {Roy}\ \emph {et~al.}(2024)\citenamefont {Roy},
  \citenamefont {Dürholt}, \citenamefont {Asche}, \citenamefont {Zipoli},\
  and\ \citenamefont {Gómez-Bombarelli}}]{roy_learning_2024}%
  \BibitemOpen
  \bibfield  {author} {\bibinfo {author} {\bibfnamefont {S.}~\bibnamefont
  {Roy}}, \bibinfo {author} {\bibfnamefont {J.~P.}\ \bibnamefont {Dürholt}},
  \bibinfo {author} {\bibfnamefont {T.~S.}\ \bibnamefont {Asche}}, \bibinfo
  {author} {\bibfnamefont {F.}~\bibnamefont {Zipoli}},\ and\ \bibinfo {author}
  {\bibfnamefont {R.}~\bibnamefont {Gómez-Bombarelli}},\ }\href
  {https://doi.org/10.1038/s41467-024-50407-9} {\bibfield  {journal} {\bibinfo
  {journal} {Nature Communications}\ }\textbf {\bibinfo {volume} {15}},\
  \bibinfo {pages} {6030} (\bibinfo {year} {2024})}\BibitemShut {NoStop}%
\bibitem [{\citenamefont {Ben~Mahmoud}, \citenamefont {Gardner},\ and\
  \citenamefont {Deringer}(2024)}]{ben_mahmoud_data_2024}%
  \BibitemOpen
  \bibfield  {author} {\bibinfo {author} {\bibfnamefont {C.}~\bibnamefont
  {Ben~Mahmoud}}, \bibinfo {author} {\bibfnamefont {J.~L.~A.}\ \bibnamefont
  {Gardner}},\ and\ \bibinfo {author} {\bibfnamefont {V.~L.}\ \bibnamefont
  {Deringer}},\ }\href {https://doi.org/10.1038/s43588-024-00636-1} {\bibfield
  {journal} {\bibinfo  {journal} {Nature Computational Science}\ }\textbf
  {\bibinfo {volume} {4}},\ \bibinfo {pages} {384} (\bibinfo {year}
  {2024})}\BibitemShut {NoStop}%
\bibitem [{\citenamefont {Unke}\ and\ \citenamefont
  {Meuwly}(2019)}]{unke_physnet_2019}%
  \BibitemOpen
  \bibfield  {author} {\bibinfo {author} {\bibfnamefont {O.~T.}\ \bibnamefont
  {Unke}}\ and\ \bibinfo {author} {\bibfnamefont {M.}~\bibnamefont {Meuwly}},\
  }\href {https://doi.org/10.1021/acs.jctc.9b00181} {\bibfield  {journal}
  {\bibinfo  {journal} {Journal of Chemical Theory and Computation}\ }\textbf
  {\bibinfo {volume} {15}},\ \bibinfo {pages} {3678} (\bibinfo {year}
  {2019})}\BibitemShut {NoStop}%
\bibitem [{\citenamefont {Ivkovi\'{c}}, \citenamefont {Jover},\ and\ \citenamefont
  {Harvey}(2024)}]{ivkovic_transfer_2024}%
  \BibitemOpen
  \bibfield  {author} {\bibinfo {author} {\bibfnamefont {\AA.}~\bibnamefont
  {Ivkovi\'{c}}}, \bibinfo {author} {\bibfnamefont {J.}~\bibnamefont {Jover}},\
  and\ \bibinfo {author} {\bibfnamefont {J.}~\bibnamefont {Harvey}},\ }\href
  {https://doi.org/10.1039/D4DD00168K} {\bibfield  {journal} {\bibinfo
  {journal} {Digital Discovery}\ }\textbf {\bibinfo {volume} {3}},\ \bibinfo
  {pages} {2242} (\bibinfo {year} {2024})},\ \bibinfo {note} {publisher:
  RSC}\BibitemShut {NoStop}%
\end{thebibliography}
%

\end{document}